\documentclass[a4paper,preprintnumbers, 11pt]{article}
\usepackage{jheppub} 
\usepackage{subfig}

\usepackage{hyperref}
\hypersetup{
    colorlinks=true,
    linkcolor=blue,
    filecolor=magenta,      
    urlcolor=blue,
    pdftitle={HI-SIGMA},
    pdfpagemode=FullScreen,
    }

\usepackage{lineno}
\usepackage{xspace}

\newcommand{\HISIGMA}{\texttt{HI-SIGMA}\xspace}

\title{Data-Driven High-Dimensional Statistical Inference with Generative Models}
\author[a]{Oz Amram}
\author[b]{Manuel Szewc}

\affiliation[a]{Fermi National Accelerator Laboratory, Batavia, IL 60510, USA}
\affiliation[b]{Department of Physics, University of Cincinnati, Cincinnati, Ohio 45221, USA and International Center for Advanced Studies (ICAS), ICIFI and ECyT-UNSAM, 25 de Mayo y Francia, (1650) San Mart\'{i}n, Buenos Aires, Argentina}
\emailAdd{oz.amram@cern.ch}
\emailAdd{szewcml@ucmail.uc.edu}

\begin{document}
\preprint{FERMILAB-PUB-25-0380-CMS-PPD}
\abstract{ 
\noindent 
Crucial to many measurements at the LHC is the use of correlated multi-dimensional information to distinguish rare processes from large backgrounds,
which is complicated by the poor modeling of many of the crucial backgrounds in Monte Carlo simulations. In this work, we introduce \HISIGMA, a method to perform unbinned high-dimensional statistical inference  with data-driven background distributions. In contradistinction to many applications of Simulation Based Inference in High Energy Physics, \HISIGMA relies on generative ML models, rather than classifiers, to learn the signal and background distributions in the high-dimensional space. These ML models allow for interpretable inference while also incorporating model errors and other sources of systematic uncertainties.
We showcase this methodology on a simplified version of a di-Higgs measurement in the $bb\gamma\gamma$ final state, where the di-photon resonance allows for background interpolation from sidebands into the signal region. 
We demonstrate that \HISIGMA provides improved sensitivity as compared to standard classifier-based methods, and that systematic uncertainties can be straightforwardly incorporated by extending methods which have been used for histogram based analyses. 

}
\maketitle

\flushbottom

\section{Introduction}
The rise of machine learning has greatly extended the physics reach of the LHC. 
Rare processes such as the Higgs decay to bottom quarks, which once were thought impossible to see at the LHC due to the overwhelming QCD background are now being measured with increasing precision \cite{CMS:2024gzs,ATLAS:2024fkg}. 
Recent projections from the ATLAS and CMS Collaborations of their sensitivity to the measurement of di-Higgs production, a sensitive of the Higgs self-coupling and a flagship measurement of the High-Luminosity LHC, have improved by a factor of ${\sim}2$ as compared to projections from only five years prior \cite{CMS:2025hfp}. 
Key to these advancements has been the usage of powerful machine learning classifiers which are capable at identifying putative signals events from backgrounds with high accuracy by using multivariate information. 

The most common usage of classifier in analysis is by placing a cut on the classifier score to reject background.
Events passing these selections are then fit in some discriminating feature, such as the mass of a resonance candidate, to extract the relevant parameter of interest. 
A more optimal, and increasingly common, alternate strategy is to use a binned classifier score as the final fit observable rather than a cut variable. 
This approach results in improved sensitivity because it makes better use of the information learned by the classifier; cuts throw away some signal events from statistical analysis, and discard information about the relative signal-versus-background likelihood of the remaining events.
The usage of binned classifier scores can be understood as a natural extension of the Matrix Element Method~\cite{Kondo:1988yd,Kondo:1991dw}, or, for the case of linear parameter dependence, the `optimal observables' framework \cite{Diehl:1993br}. 
The difference being that the Matrix Element Method uses first-principles parton-level likelihood ratios in conjunction with smearing methods to approximate non-perturbative and detector effects, a trained binary classifier learns the likelihood ratio directly from the reconstructed observables. 

Despite the power of these approaches, they also have challenges.
Most importantly, it can be difficult to estimate the background in a classifier score fit analysis because multivariate correlations among all classification features must be accounted for.
For many analyses, known mismodelings of backgrounds in simulation necessitate data-driven background estimates, which can be difficult in the multivariate case. 
Binned classifier score fits can also suffer from a lack of interpretability; the final fit observable is no longer physical and represents a collapse of a complex multivariate space to a single observable. 
This obscures the intuition of analyzers and make it difficult to spot potential mismodelings.
Such questions of interpretability may come to the forefront should any such analysis report statistical evidence of beyond standard model physics. 

Additionally, the use of binned classifiers are close to optimal only for analyses of a single parameter of interest with linear parameter dependence\cite{Brehmer:2018eca}, and therefore will require undesirable tradeoffs in the case of signal inference or multi-parameter extraction, such as measurements constraining multiple effective field theory operators. 
A class of analysis techniques, called \textit{simulation based inference} (SBI) has been proposed to optimally perform inference in this case \cite{Brehmer:2018kdj, Brehmer:2018eca, Cranmer_sbi_review}. 

Nearly all SBI proposals in particle physics are based on classifiers, making use of the property that appropriately trained neural network classifiers learn the likelihood ratio between their target classes \cite{Cranmer:2015bka}. 
These SBI methods therefore train classifiers on simulation to learn the likelihood ratio between different hypotheses under consideration, and directly use the learned likelihood ratios to perform statistical inference \cite{Ghosh_jrjc_proc, GomezAmbrosio:2022mpm, Bahl:2021dnc, Barrue:2023ysk, Schofbeck:2024zjo, Chai:2024zyl, nsbi_dihiggs, Benato:2025rgo}. 
The first application of this technique to a real LHC analysis has recently been performed by the ATLAS Collaboration \cite{ATLAS_SBI_methods, ATLAS_SBI_measurement}.
SBI methods allow for optimal multivariate inference, but have a heavy reliance on simulation, which hinders their applicability to final states which are known to require data-driven modeling. 

In this work we introduce a new approach, \textit{High Dimensional Statistical Inference with Generative Model AI} (\HISIGMA), which can perform optimal multivariate statistical inference with data-driven background modeling. 
We propose to use machine learning models to learn the multi-dimensional probability densities of signal and background, and perform statistical inference directly on these learned probability densities, rather than relying on classifiers.
We consider a resonance analysis, in which the signal is known to be localized in some signal region (SR), with sidebands on either side dominated by background events. 
Borrowing from techniques previously employed in anomaly detection \cite{Hallin:2021wme}, we use a generative model to learn the conditional multi-dimensional probability density of the background from the sideband regions, and interpolate it into the signal region.
The background probability density in the single resonant variable is estimated with a standard parametric functional form. 
These two pieces and then combined for a full estimate of the multi-dimensional background probability density. 
The multi-dimensional signal probability density is learned from Monte Carlo simulations. 
These two densities are then used to construct the likelihood function of the signal plus background hypothesis, in an analogous fashion to standard histogram-based template analyses. 
A crucial challenge in employing ML-based density estimators within statistical inference is the appropriate assessment of uncertainties to the learned density estimates. 
We show how techniques which are standard practice in uncertainty quantification in histogram-template-based density estimates can be extended to quantify the uncertainty 
on these ML-based density estimators.

We demonstrate our approach on an example measurement of di-Higgs production in the $bb\gamma\gamma$ final state. 
We show that \HISIGMA can successfully provide a multi-dimensional data-driven background estimate in such analysis, and offer sensitivity gains as compared to commonly employed classifier-based analysis strategies. 
We also demonstrate how the results of our methodology are directly interpretable, as modeling of the data by the signal plus background models can be readily inspected to check for mismodelings.

As this work was being finalized, Ref. \cite{Cheng:2025ewj} appeared, which extends Refs. \cite{Das:2023bcj,Cheng:2024yig}, and uses generative models for data-driven high-dimensional statistical inference on resonances in a similar fashion to this manuscript. 
However, Ref. \cite{Cheng:2025ewj} focuses on an anomaly detection application where the signal is unknown or partially known, whereas we focus on targeted searches, or measurements of a known signal, where more careful uncertainty quantification is required.  Additionally, the two approaches considered in Ref. \cite{Cheng:2025ewj} construct their likelihoods for statistical inference differently than \HISIGMA. In particular, neither includes any information from the resonance variable as part of the inference. We therefore view the two works as complementary.

This paper is organized as follows. In Section \ref{sec:method}, we outline the \HISIGMA method and in Section \ref{sec:sys} we discuss the incorporation of systematic uncertainties. 
We then detail an application of \HISIGMA to a di-Higgs measurement in Section \ref{sec:dihiggs} and show the results in Section \ref{sec:results}.
We discuss the various pros and cons of the \HISIGMA method as compared to commonly employed analysis strategies in Section \ref{sec:discussion}, and conclude in Section \ref{sec:conclusions}.

\section{The \HISIGMA Method}
\label{sec:method}

\HISIGMA is based on using machine learning based generative models to learn multivariate density estimates and using them to perform inference.
This approach can be considered a machine-learning-based extension of the statistical techniques that are already ubiquitously used in particle physics analyses.
Nearly all statistical analyses at colliders rely on an estimate of the probability density of signal and background over a set of features. 
These density estimates are often either estimated with histograms using Monte Carlo (MC) simulations, or based on fits to ad-hoc parametric functional forms.
Typically such approaches are limited to a very low-dimensional feature space, most often just one dimension.
Due to the curse of dimensionality, the amount of MC samples required for reliable density estimation with histograms grows exponentially with the number of dimensions, rendering it infeasible for large dimensionalities.
Parametric forms are an alternate approach, but must be hand-tuned and can have significant challenges in the presence of correlations between the variables in the  multi-dimensional case.
Recent advances in machine learning have made multi-dimensional density estimates much more tractable than previously possible, enabling their use in statistical analysis.

While these ML density estimators could be trained on simulation, their interpolation capabilities offer the opportunity to train them directly on data in appropriate control regions.
We consider the prototypical case of a resonance analysis, where the signal is known to be localized in some resonance variable, denoted $m$. 
The mass can be used to define a signal region centered around a particular mass hypothesis $m_{0}$, with the surrounding regions defining sideband control regions.
We assume that the background probability distribution varies smoothly as a function of the resonance variable, such that the background distribution in the SR can be estimated from data in the sidebands. 
We can then train a generative model in this sideband region to learn the background probability density conditional on $m$.
This conditional density can then can be interpolated into the SR and combined 
with an estimate of the background $m$ distribution, modeled with a standard parametric function, to model the full density. 
This data-driven background probability density allows then allows for high-dimensional statistical inference. 

Similar data-driven background estimation strategies for resonance searches have been extensively studied for use in anomaly detection \cite{Hallin:2021wme, Raine:2022hht, Golling:2022nkl, Hallin:2022eoq, Sengupta:2023xqy, Das:2023bcj}.
However, in the anomaly detection use case, the generative model is typically sampled from, in order to generate synthetic background events.  
A classifier is then trained to distinguish between these synthetic events and the true events in the signal region to identify any anomalous events from a localized resonance. 
In our case, we do not use the generative capabilities of the background model, but instead use its density estimation capabilities for statistical inference.

Recently, Ref. \cite{Sluijter:2025isc} compared generative and classifier-based approaches to SBI in a particle physics context and performed coverage studies.
On both a toy Gaussian example and a mock $H\to \tau\tau$ measurement, it found both approaches to have proper coverage, but the generative approach led to slightly larger uncertainties. 
However, Ref. \cite{Sluijter:2025isc} trained the generative models using only simulation, and did not consider the data-driven estimation of the background density featured in the \HISIGMA approach.
Also, due to the numerical requirements of their large suite of pseudoexperiments, Ref. \cite{Sluijter:2025isc} performed its studies on only a two-dimensional feature set, rather than the 4+1 dimensional example we showcase in our $bb\gamma\gamma$ example. As we detail below, we find that the performance of \HISIGMA is dependent on the dimensionality of feature space. It can more difficult to accurately model the probability density of higher dimensional feature spaces, but they offer potentially more information to perform inference with.

\subsection{Formalism}

A standard extended likelihood function used in particle physics to extract a signal strength in presence of a single background is given by~\cite{Cranmer:2014lly}:

\begin{equation}
    \mathcal{L} = \mathrm{Pois}(N |s+b) \cdot \prod_i^N \left(\frac{b}{s+b} \cdot P_b(\vec{x_i}) + \frac{s}{s+b} \cdot P_s(\vec{x_i})\right)
\end{equation}
where $\vec{x_i}$ is the multi-dimensional set of variables describing events $i$, $P_s$ ($P_b$) is the normalized probability density of signal (background), $s$ ($b$) are free parameters describing the rate of the signal and background processes and $\mathrm{Pois}(N| s+b)$ is the Poisson probability mass function of observing $N$ events given the expected rate $s+b$. 

The signal and background probability densities are generally not known exactly, and usually have a set of nuisance parameters $\nu$ which encode their uncertainties. The same goes for the background rate $b$.
In LHC analyses it has become standard to use the profile likelihood ratio~\cite{Cranmer:2014lly}
as the test statistic in a hypothesis test:

\begin{equation}
    \label{eq:LR}
    t_{s} = -2 \ln \frac{\mathcal{L}(\vec{x}|s,\hat{\hat{\nu}})}{\mathcal{L}(\vec{x} | \hat{s},\hat{\nu})},
\end{equation}
where $\hat{s}$ and $\hat{\nu}$ are the values of the parameters that maximize the likelihood and $\hat{\hat{\nu}}$ are the values of the nuisance parameters that maximize the likelihood for fixed $s$. The naive discovery significance can be found by computing $Z_{0}=\sqrt{t_{0}}$. Our approach does not deviate from this profile likelihood formalism. 
In \HISIGMA, $P_s(\vec{x})$ and $P_b(\vec{x})$ consist of a combination of ML-based density estimation models and parametric functional forms that allow for a data-driven background estimation.  These models are then used in a profile likelihood fit to extract best fit values of the relevant parameters and test for the presence of a signal.

To facilitate a data-driven estimate of the background density, we factorize both the signal and background probability densities into a conditional densities based on the resonance variable $m$: 

\begin{equation}
    \label{eq:likelihood_fac}
P_{k}(\vec{x}) = P_{k}(\vec{x}' |m)P_{k}(m)
\end{equation}
where $\vec{x}'$ denotes all the variables describing the event besides $m$.
The full likelihood is now given by:
\begin{equation}
    \label{eq:likelihood_full}
    \mathcal{L}(s,\nu) = \mathrm{Pois}(N | s+b(\nu)) \cdot \prod_i^N \left(\frac{b(\nu)}{s+ b(\nu)} \cdot P_b(\vec{x_i}' |m_i,\nu)P_b(m_i|\nu) + \frac{s}{s + b(\nu)} \cdot P_s(\vec{x_i}' |m_i,\nu)P_s(m_i|\nu)\right).
\end{equation} 
where $P_{k}(m)$ is a one-dimensional the probability density of the resonance variable.
This one-dimensional background $m$ density can be straightforwardly estimated from the data using the well-tested parametric functional forms used in countless collider analyses.\footnote{Alternatively, Gaussian process regression could be used if a suitable parametric form cannot be found\cite{Frate:2017mai,Gandrakota:2022wyl, Barr:2025lba}.}

The probability density of the multi-dimensional feature space conditional on $m$, $P_{k}(\vec{x}' |m)$, is learned with a generative model for both signal and background. 
For the background, the model is trained in the sideband regions on both sides of the resonance.
Conditioning on $m$, allows the learned density to be interpolated into the signal region. 
To perform inference, we also need an estimate of $P_{s}(\vec{x}'|m)$. 
In our setup, we assume the signal is known, and well described by Monte Carlo simulations. 
We therefore estimate $P_{s}(\vec{x}'|m)$ by training our generative model on simulated signal events. 
We also  estimate $P_{s}(m)$ from fits to simulated events with a standard functional form (which is in this case a Crystal Ball shape).
Note that simulations could also be used to train a generative model for any minor background not sufficiently abundant in the sidebands, such as a resonant peaking background. 

The proposed pipeline for the \HISIGMA is thus as follows:
\begin{enumerate}
    \item Define a signal region SR and sidebands SB based on the resonant variable, $m$.
    \item Estimate $P_{b}(\vec{x}'|m)$ in the SBs using a generative model of choice.
    \item Estimate $P_{s}(\vec{x}'|m)$ from MC simulations using a generative model of choice.
    \item Estimate $P_{s}(m|\nu_{s})$ from MC simulations using an appropriate parametric form.
    \item Determine an appropriate parametric function for the background mass density, $P_{b}(m|\nu_{b})$. 
    \item Use these four pieces to construct the full likelihood of the data. Perform a maximum likelihood fit to the data.
    \item Statistical quantities, such as the signal significance, and measurements of the signal strength, can then be assessed using the standard profile likelihood formalism. 
\end{enumerate}

 Although we consider only a simple one parameter measurement for this first study, the formalism can be easily extended to inference with multiple parameters of interest $\vec{\theta}$. Often in particle physics the likelihood has the form of a mixture model where the parameters of interest control the relative contributions of the different processes. This means the conditional likelihood of the different processes can be learned separately and independently of $\vec{\theta}$. 
 
 If the dependence of the likelihood in terms of the different parameters is not known a priori, MC simulations need to be generated for different values of $\vec{\theta}$ and the generative model can be trained to learn the likelihood conditional on $\vec{\theta}$, $P(\vec{x}'|m,\vec{\theta})$. Since we are learning the conditional likelihood, as in most SBI methods, the particular choice of prior distribution over parameters of interest is not important as long as the relevant region in parameter space is sufficiently covered. Having estimated  $P_{s}(\vec{x}'|m,\vec{\theta})$ and $P_{s}(m|\nu_s,\vec{\theta})$, we can perform multi-dimensional fits on $\vec{\theta}$ using any of the standard fitting tools. Thus, even if the problem becomes numerically more challenging, it is not conceptually different.
 
\section{Incorporating systematics}
\label{sec:sys}
Each piece of our likelihood function will have systematic uncertainties which need to be incorporated for reliable inference.  
Uncertainties on the parametric shape for the signal and background mass distributions can be handled in typical fashion. The background shape nuisance parameters are left freely floating in the fit, since they are constrained by the sideband data, and profiled over; while signal shape parameters and uncertainties are usually determined from simulation. 
Uncertainties affecting the multi-dimensional feature density distributions are more important and require novel strategies. 

The increased information available when fitting in high dimensions can break degeneracies between shape uncertainties and the parameter of interest, reducing the impact of systematic uncertainties.
However, this increased precision requires a corresponding increase in the rigor of the systematic uncertainty parameterization to ensure proper analysis coverage. 

We consider two different types of uncertainty affecting the learned densities.

\subsection{Uncertainty from Finite Training Data}
\label{subsec:finite_sys}

The first uncertainties we consider are statistical uncertainties arising from finite events, either in data sideband regions or Monte Carlo simulations, available for training the generative models. 
To account for these uncertainties we bootstrap our set of training events into sub-samples, and train a separate generative model for each sub-sample.
Statistical inference can then be performed with these separate bootstrapped models, and their likelihood curves combined in some fashion to determine the overall best fit and uncertainty.
One option to combine the different likelihood curves would be to use discrete profiling~\cite{Dauncey:2014xga}, which has been used in high energy physics to account for the choice of parametric functional form.
In discrete profiling, for each value of the parameter of interest (\textit{p.o.i.}) the overall log-likelihood is defined as the minimum of the log-likelihoods of the considered models.
This allows the data to `constrain' the discrete nuisance defining the choice of model; if the data prefers one model such that its log-likelihood is less than all others within the $\pm 1 \sigma$ confidence interval of the \textit{p.o.i.}, then discrete profiling will not increase the size of the confidence interval. 
In the case of alternate functional forms with low numbers of parameters, such a preference may be realistic, however in the case of profiling over different generative models a more conservative approach may be warranted.

In the \HISIGMA application of this method, the models being profiled over each have the same `functional form' but have a very large number of free parameters. 
Neural networks with large numbers of parameters are also known to essentially never reach their global minima, which is nearly always an overfitted model. 
Different models may therefore end up in quite distinct minima. 
However, when fitting large datasets, of say $O(10^4)$ events, a $O(10^{-3})$ difference in the log-likelihood of the different models will result in a $O(10)$ shift in the best fit minima, meaning discrete profiling will depend only on the `best' model within the $\pm 3 \sigma$ confidence interval. 
We therefore propose a more conservative approach, in which we define the final result as an envelope of the $\Delta NLL$ curves of the different bootstrapped models. 

To reduce the impact of outlier models which fit the data poorly, we require that the log-likelihood difference between the fit performed with a given model and one performed the `best' model (the fit with the maximum likelihood) be less than 50, to be included in the envelope. The threshold of 50 was not optimized. It represents a ${\sim}7\sigma$ preference of the data for one model versus the other. Empirically, this requirement allows the rejection of extreme outliers while preserving the conservative approach of the uncertainty.

Thus, if we have $D_{b}$ bootstrapped SB samples for background estimation that meet our criterion and $D_{s}$ bootstrapped MC samples for signal estimation, we have $d=1,\dots,D_{b}\times D_{s}$ profiled likelihoods for each $s$, $\mathcal{L}_{d}(s,\hat{\hat{\nu}})$, and we obtain the confidence intervals by combining the ``minimum substracted models'' into the modified test statistic
\begin{equation}
    \label{eq:bootstrapped_LR}
    t_{s}=\min_{d}\{- 2 \ln \frac{\mathcal{L}_{d}(s,\hat{\hat{\nu}})}{\max_{s}\mathcal{L}_{d}(s,\hat{\hat{\nu}}(s))}\}
\end{equation}
which results in the $N\sigma$ confidence interval being defined as
\begin{equation}
    [\min_{d} (\hat{s}_d-N\sigma_d), \;\max_{d} (\hat{s}_d+N\sigma_d)]
\end{equation}
for the bootstrapped models ($d$), best fit values $\hat{s}_d$ and uncertainties $\sigma_d$. 
The overall best fit value, $\hat{s}$, is taken to be the mean of the best fit values of the different bootstrapped models.

Note, in our implementation in Section \ref{sec:dihiggs}, for improved stability we take each bootstrapped model to be an ensemble of ten generative models trained on the same dataset with different initial seeds. 
To further prevent overfitting, we divide the boostrapped sample in training and validation samples, and stop the training if the validation loss has not improved after a certain number of epochs. 
The final model is taken as the model from the epoch with the lowest validation loss. The log-probabilities of each individual model are averaged to produce the final density estimate. We find that this procedure of early stopping and ensembling reduces model variability between the different bootstraps.

\subsection{Shape uncertainties}
\label{subsec:shape_sys}

An additional important uncertainty to consider are variations of the feature distributions from known sources of systematic uncertainty. 
In histogram-based analyses, these are accounted for by producing an `up' and `down' variation of the relevant distribution, corresponding to $\pm1\sigma$ variations of the systematic uncertainty.
A nuisance parameter, $\nu$, is included into the fit which captures these systematic variations. 
We define $\nu=0$ as the nominal distribution, whereas $\nu=\pm1$ represents the up/down variations. 
A Gaussian constraint is added to the likelihood to penalize variations away from $\nu=0$.
For intermediate values of $\nu$, an interpolation scheme is used to produce an estimate of the density in between the nominal and up/down variations (see Ref. \cite{Cranmer:2014lly} for a review). 

For an extension of this method to our unbinned density estimates we follow a a similar strategy to Ref.~\cite{ATLAS_SBI_methods}. 
For each systematic uncertainty, we train a separate model for datasets drawn from the relevant up/down variations of the systematic. 
To maintain consistency with the method of Sec. \ref{subsec:finite_sys}, this is done separately for each boostrapped sample. 
We then use a parametric interpolation to define the density estimate as a function of the shape systematic $\nu$
\begin{equation}
    \label{eq:model_with_syst}
    \begin{split}
    P_{k}(\vec{x}'|m,\nu)&=P_{k}(\vec{x}'|m,\nu=0)\frac{P_{k}(\vec{x}'|m,\nu)}{P_{k}(\vec{x}'|m,\nu=0)}\\
    &=P_{k}(\vec{x}'|m,\nu=0)w_{k}(\vec{x}',m,\nu)
    \end{split}
\end{equation}
where
\begin{equation}
    \label{eq:weights_for_syst}
    w_{k}(\vec{x}',m,\nu)=
    \begin{cases} 
    \left(\frac{P_{k}(\vec{x}'|m,\nu=1)}{P_{k}(\vec{x}'|m,\nu=0)}\right)^{\nu} & \text{ if $\nu > 1$}\\
    1+\sum_{i=1}^{6}c_{ki}(\vec{x}',m)\nu^{i}  & \text{ if $|\nu|\leq 1$}\\
    \left(\frac{P_{k}(\vec{x}'|m,\nu=-1)}{P_{k}(\vec{x}'|m,\nu=0)}\right)^{-\nu}& \text{ if $\nu < -1$}\\
    \end{cases}
\end{equation}
and $P_{k}(\vec{x}'|m,\nu=0)$ ($P_{k}(\vec{x}'|m,\nu=\pm 1)$) is the nominal ($\pm 1 \sigma$) model as obtained from averaging over the ensemble of models obtained with the nominal ($\pm 1\sigma $) systematic variation of the MC samples, and $c_{ki}(\vec{x}',m)$ are chosen so that the function and its first two derivatives are continuous at $\nu = \pm 1$. The continuity condition defines a set of six linear equations which can be used to solve explicitly for the interpolation coefficients $c_{ki}(\vec{x}',m)$:
\begin{equation}
    \begin{split}
        1+\sum_{i=1}^{6}c_{ki}(\vec{x}',m)(\pm1)^{i}&=\frac{P_{k}(\vec{x}'|m,\nu=\pm1)}{P_{k}(\vec{x}'|m,\nu=0)}\,, \nonumber\\
        \sum_{i=1}^{6}c_{ki}(\vec{x}',m)i(\pm1)^{i-1}&=\pm\frac{P_{k}(\vec{x}'|m,\nu=\pm1)}{P_{k}(\vec{x}'|m,\nu=0)}\ln \frac{P_{k}(\vec{x}'|m,\nu=\pm1)}{P_{k}(\vec{x}'|m,\nu=0)}\,, \nonumber\\
        \sum_{i=2}^{6}c_{ki}(\vec{x}',m)i(i-1)(\pm1)^{i-2}&=\frac{P_{k}(\vec{x}'|m,\nu=\pm1)}{P_{k}(\vec{x}'|m,\nu=0)}\left(\ln \frac{P_{k}(\vec{x}'|m,\nu=\pm1)}{P_{k}(\vec{x}'|m,\nu=0)}\right)^{2}\,. \nonumber
    \end{split}
\end{equation}
Note that these interpolation coefficients,  $c_{ki}(\vec{x}',m)$, are determined uniquely for each data point, because all points must have continuous density estimates.
This an extension of the binned case in which the coefficients are determined separately for each bin. 
The weights are incorporated into the likelihood model and $\nu$ is profiled over. This approximation allows to include in principle several systematic sources which are assumed to be independent while keeping the number of trained models relatively under control.

Although the flexibility of Eq.\ref{eq:simpler_weights_for_syst} permits the modeling of complex shape uncertainties, the parameterization can be too flexible in certain cases. In particular, we empirically found that the coefficients overfit what turns out to be a simpler dependence on $\nu$ for the example considered in Section \ref{sec:dihiggs}. We thus consider a less complex parameterization of the weights,
\begin{equation}
    \label{eq:simpler_weights_for_syst}
    w_{k}(\vec{x}',m,\nu)=1+\sum_{i=1}^{2}c_{ki}(\vec{x}',m)\nu^{i}
\end{equation}
with no piecewise splitting and only two learnable coefficients, which can be fitted by evaluating the weights at $\nu= \pm 1$. We empirically validate this choice through a closure test, detailed in Section \ref{sec:dihiggs}. Although in this work we do not consider the full form of Eq.~\ref{eq:weights_for_syst}, it is useful to understand \ref{eq:simpler_weights_for_syst} as its simplification and we expect the former to be necessary for more complex shape uncertainties.

\section{Example Application : Di-Higgs Measurement}
\label{sec:dihiggs}
To showcase the proposed methodology we consider a simplified version of an important LHC analysis: the search for di-Higgs production in the $bb\gamma\gamma$ final state. 
Measuring di-Higgs signals at the LHC and HL-LHC tests the electroweak sector of the SM by constraining the Higgs potential~\cite{Dawson:2022zbb}. Even more, because the SM prediction is suppressed due to an accidental cancellation between diagrams that is only present for the SM potential, any excess could point towards BSM contributions~\cite{Bishara:2016kjn}. 
Due to the low expected number of events and complicated final state topologies which require data-driven backgrounds, measuring the SM di-Higgs is challenging without multivariate techniques.
Projections in the sensitivity of ATLAS and CMS to di-Higgs with the full HL-LHC dataset have improved by a factor of $\sim 2$ in the last several years due to improved analysis methodologies, including the usage of multivariate methods \cite{CMS:2025hfp}. 
The $bb\gamma\gamma$ final state features a narrow mass peak in the $m_{\gamma\gamma}$ distribution, but the use of multivariate information, exploiting correlations between kinematics of the four final state objects in the event, is still required for background suppression. 
Data-driven background estimation is also required, as simulation modeling of the background in this final state is not of sufficiently high quality. Previous efforts for fully data-driven density estimation have been proposed, see e.g. Ref.~\cite{Alvarez:2022kfr}, but rely on specific parametric functional forms which limits their scalability to incorporate necessary features for background suppression.
Thus, the analysis is an ideal demonstration of the \HISIGMA approach, which uses fully unbinned multivariate analysis with data-driven backgrounds.

Previous work has explored the use of classifier-based SBI methods on this final state \cite{nsbi_dihiggs}, demonstrating gains in constraining effective field theory (EFT) operators as compared to classical methods. 
However as an exploratory work, this previous paper did not address how such a classifier-based SBI approach, which relies on learning the likelihood ratio using simulated backgrounds, could be employed in an analysis requiring a data-driven background modeling. 
Our work addresses this challenge, demonstrating how a data-driven background estimation can be used in multivariate statistical inference. 
Though \HISIGMA is in principle able to optimally perform multi-parameter inference as well, and thus should give sensitivity gains in the simultaneous extraction of multiple EFT operators as compared to traditional methods, we leave a such a study of this capability, and a comparison of the sensitivity of the approach of Ref. \cite{nsbi_dihiggs}, for future work.

To benchmark the \HISIGMA strategy, we simulate SM di-Higgs events decaying to $bb\gamma\gamma$ and its main irreducible non-resonant $bb\gamma\gamma$ background. 
There is also a significant background of $jj\gamma\gamma$ where jets ($j$) originating from light quarks or gluons which are mistakenly identified as b-jets, but because the kinematic shapes of this background are very similar to the $bb\gamma\gamma$ background we do not include it as a separate component.
Additionally, there is also the sub-dominant but crucial background from single-Higgs production, mainly in association with a vector boson $V$ through the $VH, V\to bb H\to\gamma \gamma$ channel, to consider. 
Such a background also has a resonant peak at $m_{\gamma\gamma} \sim 125$ GeV and is not learnable from data sideband regions in $m_{\gamma \gamma}$. 
In an actual analysis, the probability distribution of such a background could be learned from simulation, similar to how this is done for the di-Higgs signal, and included as an additional component in the fit.
However in this proof-of-concept work we neglect this background for simplicity. 

The signal and background processes are simulated with \verb|MadGraph_aMC@NLO|~\cite{Alwall:2014hca} v3.5.7 at a center of mass energy of 14~TeV. 
Hadronization was done with \verb|Pythia8|~\cite{Sjostrand:2014zea,Bierlich:2022pfr}. 
All samples are generated at leading order using the NNPDF2.3 LO parton distribution function~\cite{Ball:2010de, NNPDF:2014otw}. Detector simulation and reconstruction is then performed using \verb|Delphes|~\cite{deFavereau:2013fsa}, using the default HL-LHC detector settings. 
Jets were clustered using the anti-$k_T$ algorithm with $R=0.4$ as implemented in \verb|FastJet|\cite{Cacciari:2011ma} via \verb|Delphes|.
For computational efficiency, a requirement was placed in the \verb|MadGraph_aMC@NLO| generation requiring $m_{bb} > 40$ GeV and $m_{\gamma\gamma} > 80$ GeV.

We select events with 2 $b$-tagged jets and two isolated photons.
The leading $b$-jet and photon are required to have $p_T > 30$ GeV and the sub-leading $p_T > 20$ GeV. 
$\gamma$'s and $b$'s are are required to have a rapidity $\eta < 2.4$. 
The highest two $p_T$ $b$-jets and $\gamma$'s are used to construct the $H_{\gamma\gamma}$ and $H_{bb}$ candidates.
The angular separation, $\Delta R_{ij}=\sqrt{(\phi_{i}-\phi_{j})^{2}+(\eta_{i}-\eta_{j})^{2}}$, between each $\gamma$-$b$ pair, is required to be $\Delta R_{\gamma b} > 1.0$, in order to suppress background processes with photon final state radiation.
We generate sufficient events such that $\sim$1M background events and $\sim$200k signal events pass these selection criteria.

We note that this number of background events is several orders of magnitude smaller than what can be expected in the HL-LHC dataset.
Such a larger background sample would increase the statistics available to train the generative model in the sidebands, likely improving its quality and decreasing variability.
However, this also means the signal injections we perform for testing our method, which consider signal strengths of S/B $\sim 10^{-2}-10^{-3}$ are all unrealistic for a true di-Higgs $bb\gamma\gamma$ analysis.
It is likely that our background models would not be accurate at very small probabilities, of say $10^{-6}$, making it difficult to extract a signal with an S/B of this magnitude.
However, one could imagine applying some additional selection criteria, perhaps using a multivariate classifier, to the data prior employing to \HISIGMA, so as to enhance the expected signal fraction to the level of sensitivity of our generative modeling capabilities.
Improvements to employed generative models could also increase the sensitivity of this method to smaller signal strengths.
We leave these explorations for future work.

We use $m_{\gamma\gamma}$ as the resonant feature since the good photon momentum resolution allows for the measurement of a sharp resonance against a smoothly decaying background. We consider 4 additional event features useful for distinguishing signal and background:
\begin{equation}
    \label{eq:xdef}
    \vec{x'}=\{p_{T}^{bb},p_{T}^{\gamma\gamma}/m_{\gamma \gamma}, \Delta R_{bb}, \Delta R_{\gamma\gamma}\}
\end{equation}
where $p_{T}^{bb}$ ($p_{T}^{\gamma\gamma}$) is the transverse momentum of the $H_{bb}$ ($H_{\gamma\gamma}$) candidate, and $\Delta R_{bb}$ ($\Delta R_{\gamma\gamma}$) is the angular distance between two b's (two $\gamma$'s).
There are other potential features which could be added to improve signal and background separation, such as the $H_{bb}$ candidate mass, and $\Delta R$ variables between the different $b-\gamma$ combinations, however we will stick to this reduced 5D feature set as a first proof of concept.

We define our signal region and sidebands as:
\begin{equation}
    \label{eq:sr_sb_def}
    \begin{split}
        \text{SR } &: m_{\gamma\gamma} \in [115, 135)\text{ GeV}\\
        \text{SB } &: m_{\gamma\gamma} \in [90,115)\cup [135,180)\text{ GeV}\\
    \end{split}
\end{equation}

\begin{figure}
    \centering
    \includegraphics[width=0.49\linewidth]{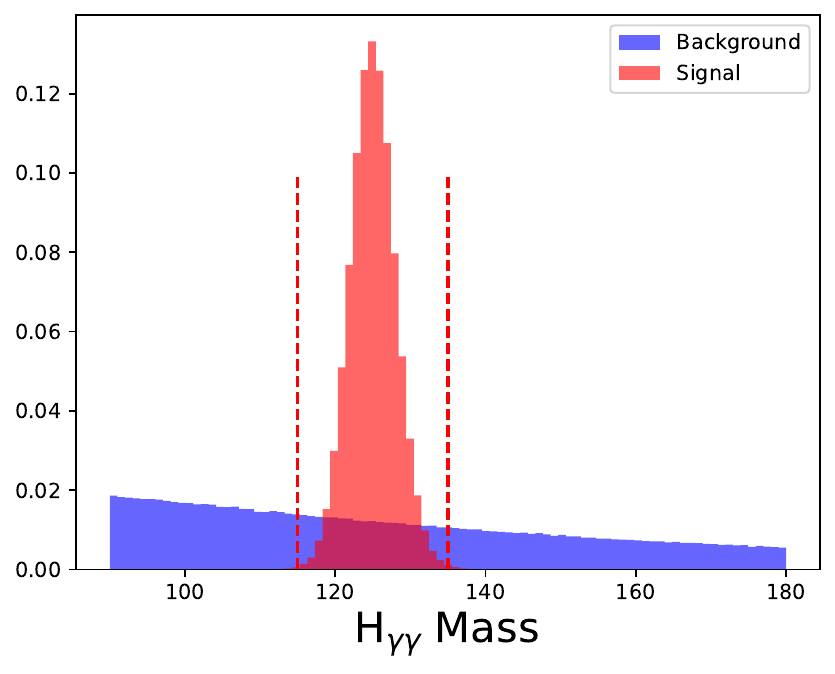} \\
    \includegraphics[width=0.49\linewidth]{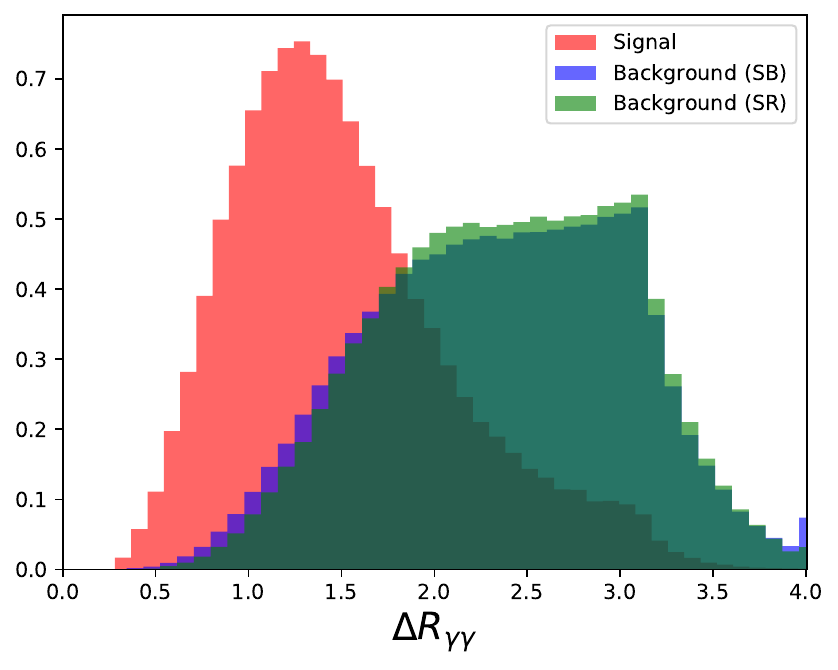}
    \includegraphics[width=0.49\linewidth]{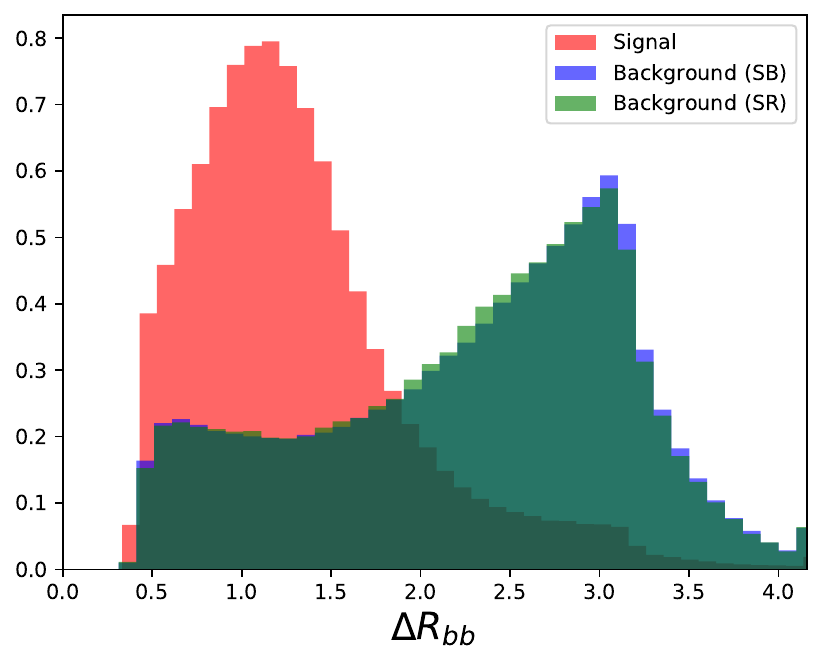}
    \includegraphics[width=0.49\linewidth]{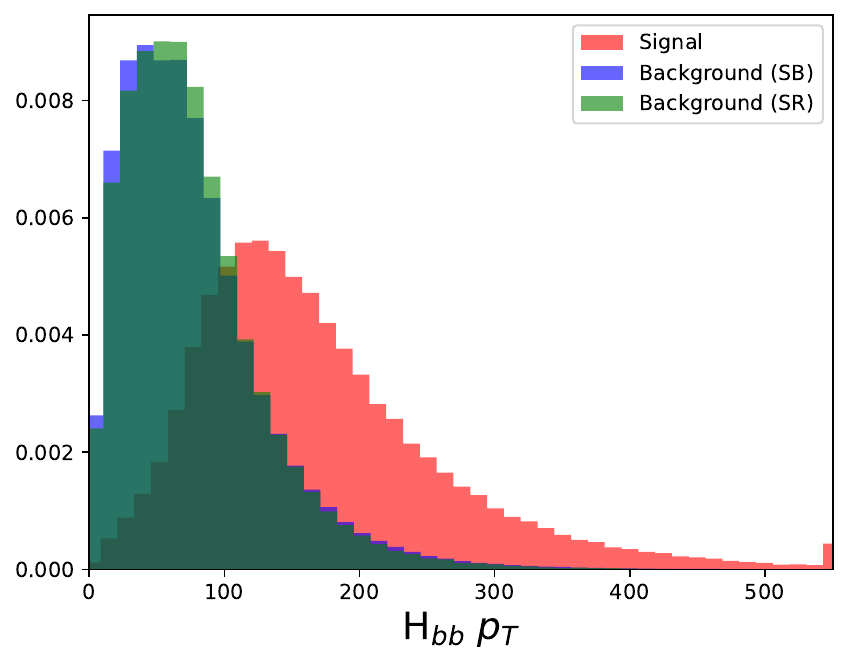}
    \includegraphics[width=0.49\linewidth]{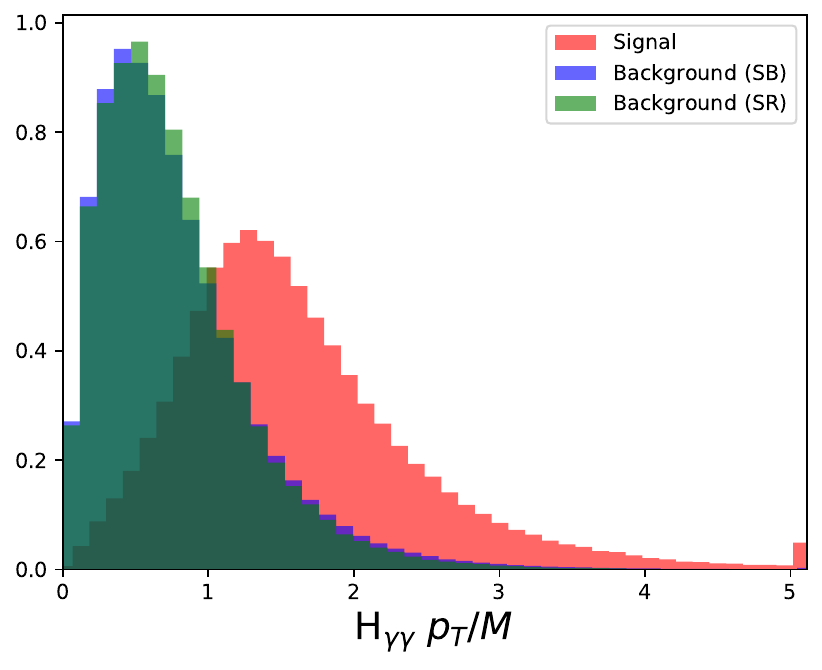}
    
    \caption{Distributions of the signal (red) and background in the sideband region (blue) and signal regions (green).
    In the top plot the dashed red lines denote the boundaries of the signal region defined by $m_{\gamma\gamma}$. 
    The last bin shows overflow entries. 
    The distributions of background events in the signal and sideband regions are seen to be similar but with visible differences, which will be learned by the interpolation aspect of the \HISIGMA approach.}
    \label{fig:feats}
\end{figure}

We use 70\% of the background and signal datasets for training and validation, and holdout the remaining 30\% for testing.  Distributions of our event features for background events in the signal region, background events in the sideband, and signal events are shown in Fig. \ref{fig:feats}.

 \subsection{Data Smearing}
Our density estimation model of choice, normalizing flows, are known to have difficulties modeling sharp features and hard boundaries in feature distributions. 
Unfortunately,  physics variables often have hard boundaries due to prior selections, such as isolation requirements leading to a minimum $\Delta R$ of 0.4 between objects, or kinematic thresholds. 
Adding a small amount of Gaussian noise to the feature can soften these hard boundaries and make it easier to model for the normalizing flow. 
However, because the trained model must be used to evaluate the density on data matching its training set, this means that the actual data being fit must be smeared as well.
This data smearing results in a loss of information and therefore results in a loss of true optimality.
However, such a smearing can also act as a regularizer, limiting analysis sensitivity to the regime where the generative model can accurately model changes in the density. 
In our testing we found the application of such smearing necessary to obtain sufficiently high quality modeling of the backgrounds. 

\begin{figure}
    \centering
    \includegraphics[width=0.35\linewidth]{figs/features/dR_bb.pdf}
    \includegraphics[width=0.35\linewidth]{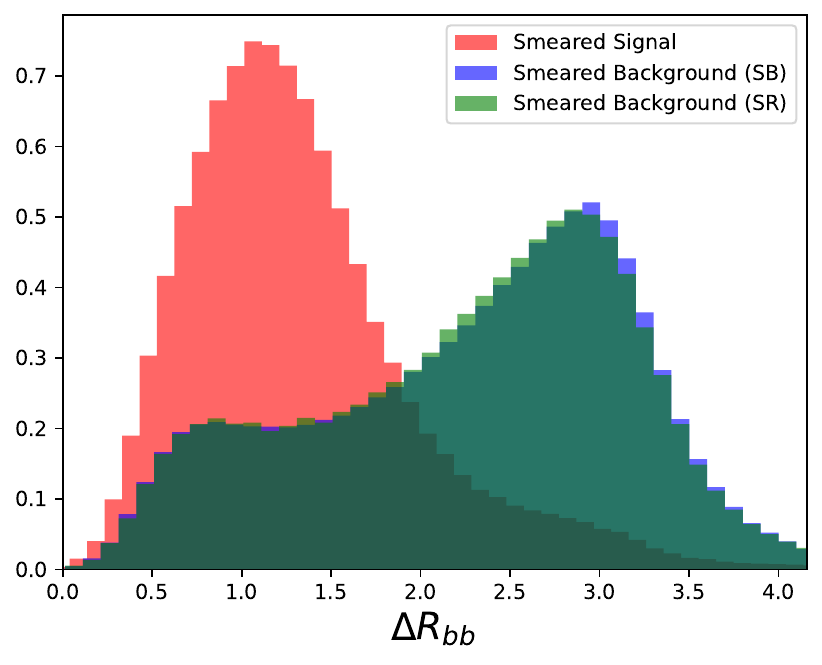}
    \caption{Distributions of the $\Delta R_{bb}$ feature before (left) and after (right) the smearing procedure. The smearing softens the sharp edge at 0.4. It incurs some information loss but does not significantly impact signal versus background discrimination. }
    \label{fig:pre_after_smearing}
\end{figure}

The amount of smearing applied to the data is therefore a very important choice, directly impacting the sensitivity of the final analysis.
In this work, we found smearing each feature by 0.2 standard deviations allowed modeling of the distributions with high accuracy while resulting in only minor sensitivity loss.
The resonance feature $m_{\gamma\gamma}$, which is not modeled by the normalizing flow, is not smeared. A visualization of one feature distribution before and after smearing is shown in Fig. \ref{fig:pre_after_smearing}.
It is possible that with an improved machine learning model, less smearing, or perhaps no smearing, would be necessary.
Alternatively perhaps a less simplistic smearing procedure, perhaps smearing some features and not others, or only for data near sharp boundaries could be employed to better mitigate sensitivity loss. 
As such considerations are very application dependent, we leave such explorations for future work. 

\subsection{Pre-processing, Model Architecture and Validation}
\label{subsec:model_architecture}
After smearing, the features are passed through other pre-processing steps before being passed to the generative model. 
These steps generally morph the distribution of each feature closer to a Gaussian distribution, making it easier to model for the normalizing flow. 
These steps are all invertible and incur no information loss, and therefore unlike the smearing do not diminish sensitivity.
First, features are shifted and scaled to the range $x \in (0,1)$.
Then a logit transformation ($\text{logit}(x)=\ln\left(\frac{x}{1-x}\right)$) is applied to convert bounded domains into unbounded ones.
The transformed features are then standard scaled to have mean zero and standard deviation of one.
To avoid outliers, features then are clipped to be within the range [-5,5].

 The generative models used to learn the signal and background conditional feature distributions, $P_{k}(\vec{x}'|m)$, are normalizing flows. Normalizing flows~\cite{Kobyzev:2021a,Rezende:2015a,Laurent:2015a} estimate a given probability distribution $P_{k}(\vec{x}'|m)$ over a feature space $\vec{x}'$ by learning an invertible mapping between $\vec{x}'$ and a latent $\vec{z}$ of the same dimension, $\vec{x}'=f(\vec{z},m)$. The probability distribution can be written explicitly in terms of the function $f$, its inverse $f^{-1}$, its gradient $\nabla_{\vec{z}}f$ and the base distribution $P(\vec{z})$:
 \begin{equation}
    P_{k}(\vec{x}'|m) = P(\vec{z}=f^{-1}(\vec{x}',m))||\det \left(\nabla_{\vec{z}}f\right)|_{\vec{z}=f^{-1}(\vec{x}',m)}||^{-1}\,.
    \label{eq:flow_prob}
 \end{equation}
$P(\vec{z})$ is chosen to be a simple base distribution that allows for sample generation and density evaluation. In this work, we consider a multivariate normal distribution $P(\vec{z})=\mathcal{N}(\vec{0},\mathbb{I})$. The function $f$ is a learnable invertible function, written in terms of Neural Networks and optimized by minimizing the negative log-likelihood derived from Eq.~\ref{eq:flow_prob}. The structure of $f$ needs to be chosen carefully to allow for evaluation not only in the forward direction ($\vec{x}'=f(\vec{z},m)$) used for sampling but also in the backward direction ($\vec{z}=f^{-1}(\vec{x}',m)$) necessary for likelihood evaluation and training. Additionally, the Jacobian of the function should be easily computable to ensure numerically efficient training and likelihood evaluation. 

In this work, we use a Masked Autoregressive Flow~\cite{papamakarios2017masked}, where $f$ consists of the composition of 6 rational quadratic spline layers~\cite{MPRQAF2021}, each with 24 bins and a hidden size of 64.
The total number of trainable parameters is 220k.
BatchNorm~\cite{batchnorm} is used in between every flow layer. The flows are implemented using the \verb|nflows| package~\cite{nflows} and are trained up to 100 epochs.
Early stopping is triggered if no improvement is seen on the validation loss after 10 epochs.
To account for training variability, five models are trained on the same dataset with different random seeds. The final model is taken as the ensemble of these 5 different models, averaging their log-probabilities to produce the final density estimate during evaluation. 

The quality of the background models was checked by sampling from them and comparing the generated samples against the true background events.
This is first done for background events in the sideband region in which the model was trained.
A classifier trained to distinguish between the real and artificial background events, using a neural network architecture as described below, produces an AUC on of $\sim 0.501$ indicating very high quality background modeling.

\begin{figure}
    \centering
    \includegraphics[width=0.49\linewidth]{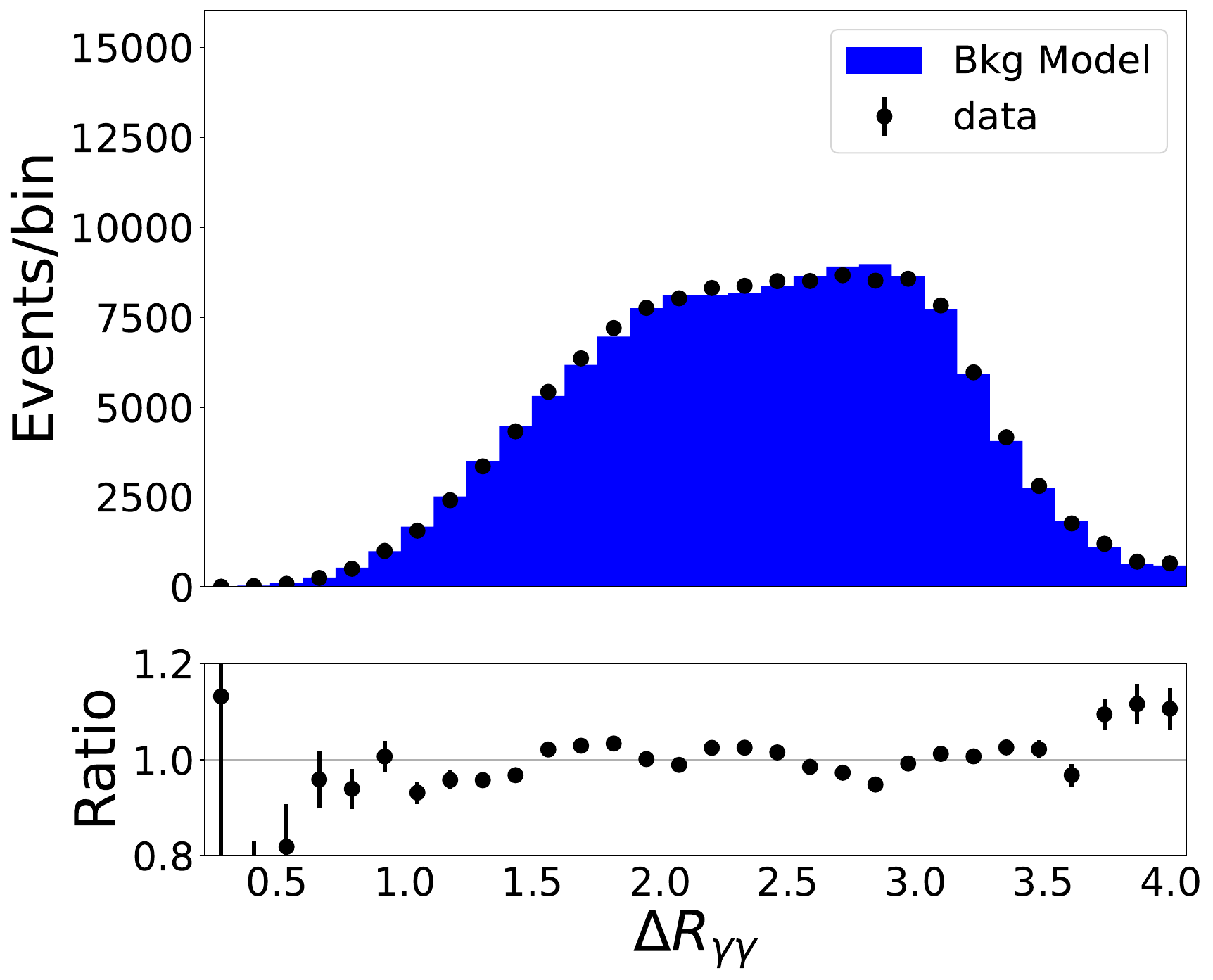}
    \includegraphics[width=0.49\linewidth]{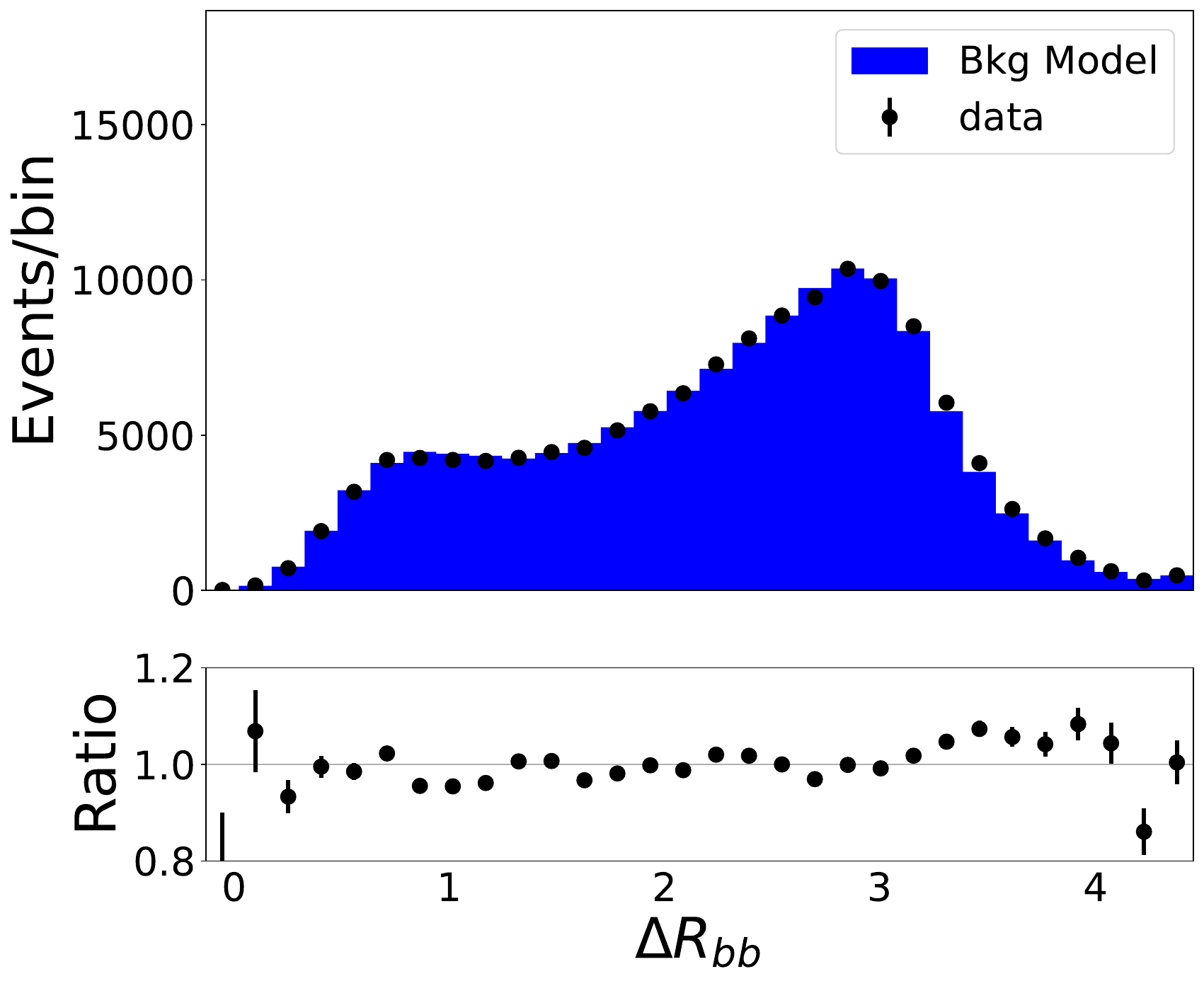}
    \includegraphics[width=0.49\linewidth]{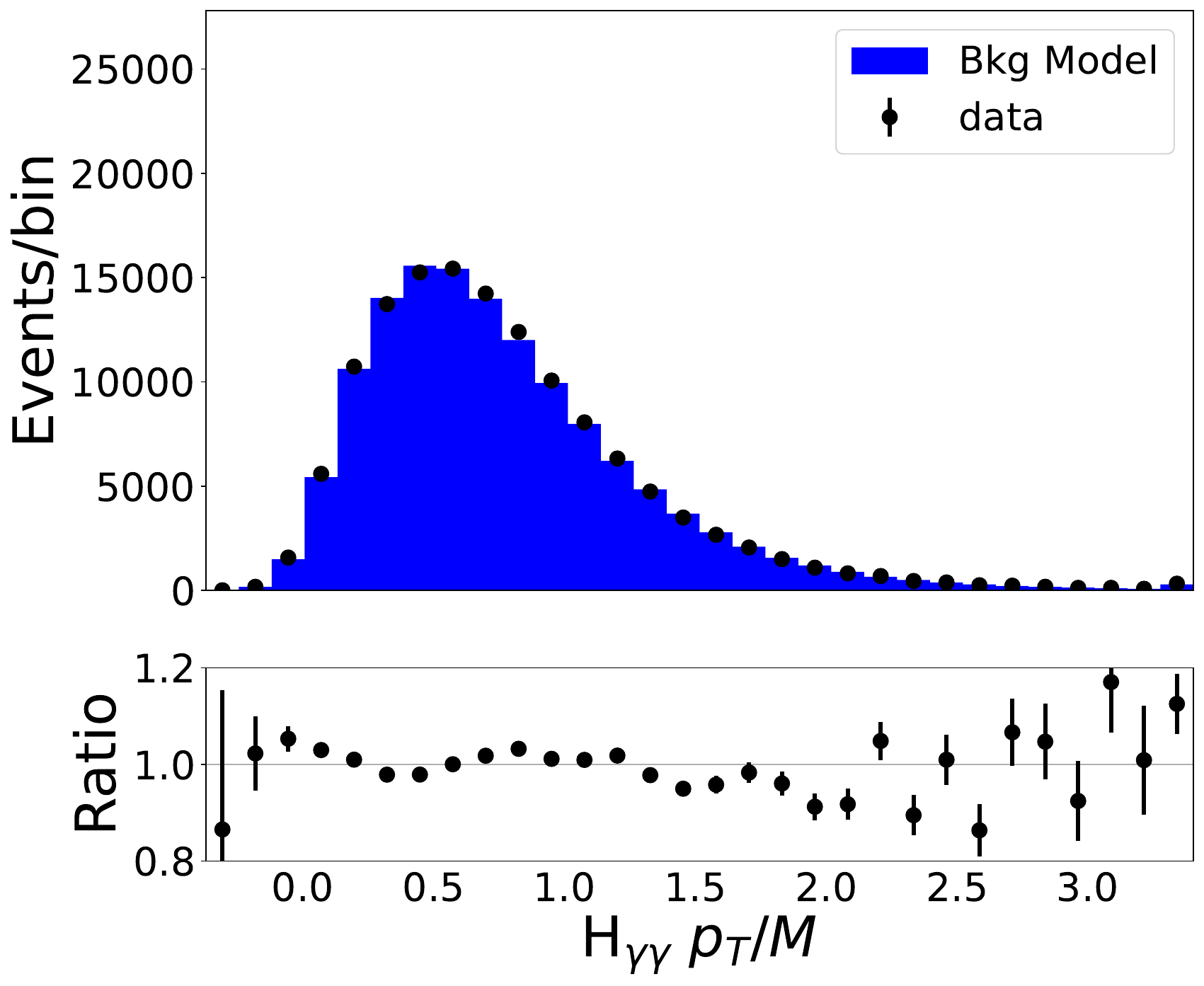}
    \includegraphics[width=0.49\linewidth]{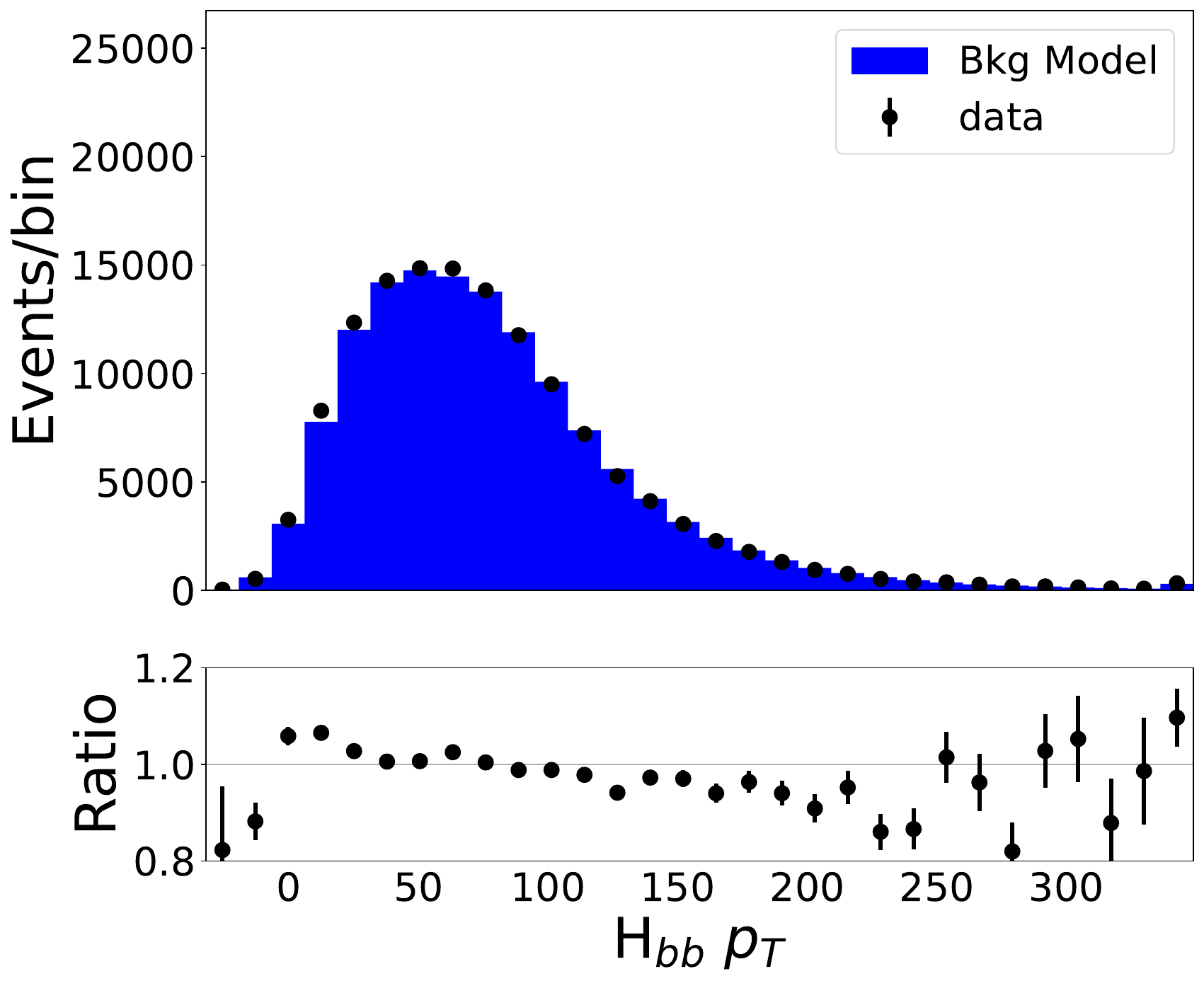}
    \caption{Events generated from a background model (single bootstrap) as compared to true events in the signal region. To spot potential minor mismodelings, the comparison is performed using samples an order of magnitude larger than those used for the inference example. Good agreement is observed overall, with only minor discrepancies visible in the tails. }
    \label{fig:training_validation}
\end{figure}

We then test the extrapolation of the background models into signal region. 
We generate events in the signal region and compare them to true background events from the signal region. 
An example comparison between true events and those sampled from a single bootstrapped model is shown in Fig.~\ref{fig:training_validation}.
Despite the model never having seen events from this region, we observe quite good modeling due to the successful interpolation.
There are some minor residual discrepancies in the tails which vary between the bootstraps.
We train a classifier to distinguish between real and generated events in this signal region and find an AUC of $\sim 0.505$.
This is slightly degraded as compared to the modeling of events in the sidebands but still quite high quality. 
We note that this validation is done using the entire training/validation set consisting of 700k events, which is approximately 10 times as many events as are used in the example fits in Section \ref{sec:results}.
Therefore the minor imperfections noted in this test may not be prohibitive for accurate parameter estimations on a smaller datasets with statistical uncertainties significantly larger than these minor density mismodelings. 
Further work would be required to properly parameterize the uncertainties on the density estimates arising from these subtle modeling imperfections.

Additionally, we note that this validation of the interpolation capabilities of the background model would not be possible in a true measurement on data, as it requires the usage of events from the signal region.
However a similar validation could be performed in data in some control region, perhaps using high $M_{\gamma \gamma}$ events above the Higgs mass. A separate training could be performed on events from this region, holding out a specific mass window from the training, and then testing the model interpolation into this region. 

\subsection{Parametric Functional Forms}
Our method also requires an estimation of the 1D probability densities of the resonance feature, $P_b(m)$ and $P_s(m)$, for which we employ well-tested parametric functional forms.
For the signal shape, we use a double Crystal Ball function function \cite{Oreglia:1980cs,Gaiser:1982yw} based on a fit to signal MC events.
The Crystal Ball shape is commonly used to model resonances and is composed of a Gaussian core and power law tails. 
For the background shape we use an exponential distribution of the form:

\begin{equation}
    P_b(\bar{m}) = p_0 e^{-p_1\bar{m} + p_2\bar{m}^2},
\end{equation}
where the $p_i$ are floating free parameters, and $\bar{m}$ is just a rescaled version of $m_{\gamma\gamma}$, scaled to be in the range [0,1]. 

\subsection{Example Shape Uncertainty}

\begin{figure}
    \centering
    \includegraphics[width=0.5\linewidth]{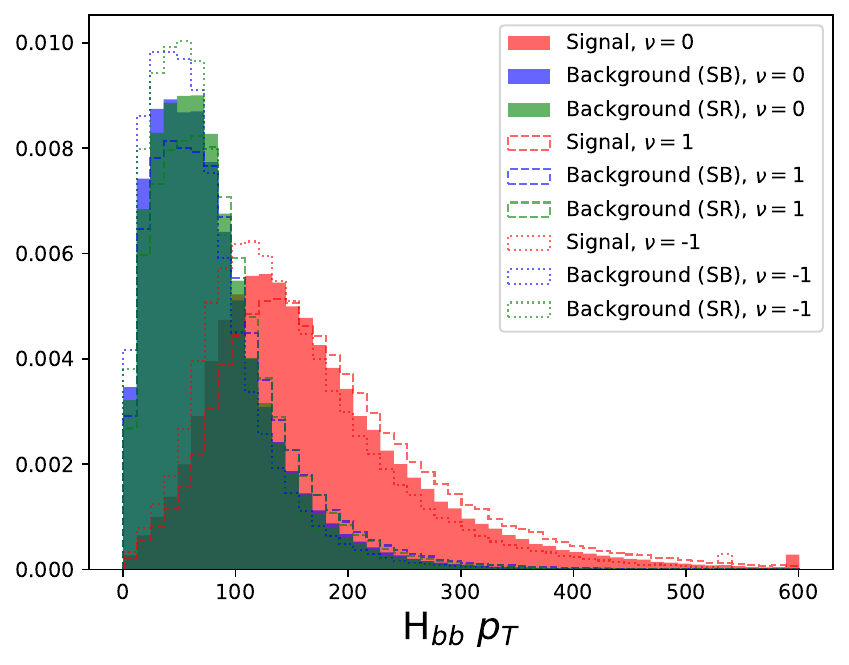}
    \caption{Distributions of the $p_{T}^{bb}$ feature for $-\sigma$, nominal and $\sigma$ variations of the shape systematic.
    The other features are assumed to be unaffected.}
    \label{fig:hbb_smearing}
\end{figure}

To deploy of \HISIGMA in any real measurement, it needs to be able to incorporate systematic uncertainties. 
To demonstrate that this is possible, we implement a toy shape systematic within this di-Higgs example following the method detailed in Sec. \ref{subsec:shape_sys}. 
We consider a $10\%$ relative uncertainty associated with the $p_{T}$ of the $H_{bb}$ candidate, such that for a given $\nu$ we re-scale $p^{bb}_{T}\to(1+0.1\nu)p^{bb}_{T}$ mimicking the effects of a jet energy scale uncertainty.
This is done for both signal and background distributions, with both shifts parameterized by a single common shared nuisance parameter. 
We show the effects of the $\pm 1 \sigma$ variations in Fig.~\ref{fig:hbb_smearing}.

For this simple uncertainty, where the $\nu$ dependence is certainly present but can be modelled with a low-complexity function, we consider the simple parameterization of the weights introduced in Eq.~\ref{eq:simpler_weights_for_syst}.
We performed a closure test to verify that the interpolation works as expected. 
Mock datasets were generated from the $\nu = 0,\pm 1$ models, and successfully fit with the full model. 
We verified that the fitted signal strength and nuisance parameter agreed well with the true values, $(\hat{s},\hat{\nu})=(s_{\mathrm{true}},\nu_{\mathrm{true}})$.

\subsection{Comparison Analysis Strategies}
\label{subsec:comparison_analysis_strategies}

To benchmark the performance of \HISIGMA we compare it to alternate analysis strategies.  The first strategy we compare to is a fit to the $m_{\gamma\gamma}$ distribution with no additional selection criteria applied. This strategy is not meant to be a realistic representation of an optimized analysis, but illustrates how much of \HISIGMA's constraining power comes from the mass information versus the multivariate feature information.

The next two analysis strategies employ DNN classifiers to discriminate signal versus background, similar to what is done in many current analyses.
The first strategy employs a `cut and fit' approach.
A DNN classifier is trained to distinguish signal versus background using the feature information, $\vec{x}'$, but not the resonance feature $m_{\gamma\gamma}$. 
A cut is placed on the classifier score to reduce the background and the $m_{\gamma\gamma}$ distribution of the remaining events is fit to extract the signal.
The cut value was chosen to maximize the signal significance, approximated as $s/\sqrt{b}$. 
Using a cut-based approach such as this is sub-optimal, as some signal events are excluded from the final analysis, and one throws away any information on signal versus background discrimination of the remaining events.
Often, multiple exclusive categories based on different classifier score thresholds defined simultaneously fit, so as to mitigate this sensitivity loss.
This strategy is employed in recent ATLAS and CMS analyses of di-Higgs $bb\gamma\gamma$ \cite{CMS_bbgg, ATLAS_bbgg, ATLAS_bbgg_run3}. 
However we stick with a single category here for simplicity. 

The third alternate analysis method uses a classifier trained using the full event information, the features and the resonance variable, to define an optimal summary statistic. A 1D histogram of the distribution of classifier scores in data is then fit to the sum of signal and background templates to extract the signal strength.
Although the classifier will learn a monotonic rescaling of the likelihood ratio, the optimal observable for signal versus background discrimination, binning will result in some loss of information.
Larger numbers of bins will mitigate this information loss and result in higher sensitivity. 
In our example we compare two choices for the number of classifier bins, 10 and 50.

Both DNN-based analyses use almost the same DNN architecture, differing only on the size of the input. The input layer of both DNNs has a BatchNorm layer~\cite{batchnorm} to scale the features appropriately. Both DNNs possess five inner layers with 32,64,64,32 and 8 neurons, respectively, and a use a and a SiLU activation function after the first layer and Leaky ReLU for the other layers. The final layer consists of a single  single neuron and a sigmoid activation function.  To avoid overfitting, a Dropout layer is added after the input and inner linear layers, with a dropout probability of $0.2$. Both models are implemented in \verb|Pytorch| and trained to minimize the Binary Cross-Entropy loss, with the same early stopping procedure as the normalizing flows but with $15$ epochs. The DNN architecture was not extensively optimized, but it was checked that increasing the size of the hidden layers did not improve performance.

This third alternate analysis requires an accurate model of the background classifier score shape.
Because the classifier score distribution has complex multivariate correlations with all the input features, estimating the background shape using data-driven methods can be difficult and methods are usually analysis-specific.
For purposes of sensitivity comparison, in this study we consider an idealized setting where the signal and background classifier score shapes are known exactly. 
This allows a comparison with the \HISIGMA sensitivity in an idealized setting in which systematic uncertainties are not assessed on either method. 
However, it should be noted that the \HISIGMA method automatically includes a data-driven background estimate, whereas for the classifier score analysis such an estimate may be quite difficult.

The fits for all analyses strategies are done with the \texttt{iminuit} \cite{iminuit} package which provides a pythonic interface to the ROOT \texttt{Minuit} minimizer \cite{James:1975dr}. 
We should note that for the \HISIGMA fits, because the machine learning model encoding the signal and background densities are fixed, the densities for each data point are computed ahead of time, avoiding computationally costly evaluations of the normalizing flow model within the fit. 
This means the \HISIGMA fits are not any more computationally intensive than the comparison methods.

\subsection{Toy Simulation Mismodeling Example}

We should note that the optimality of the DNN-based analyses depends on the accuracy of the background simulations used for training, which is the very issue that motivates \HISIGMA. 
Even if a subsequent data-driven background estimate is performed, the performance of the DNN classifier may be sub-optimal if the background simulations used for its training are inaccurate. 
Since \HISIGMA learns the background likelihood directly from data, it does not suffer from this limitation and may offer an advantage over the DNN in this case.

To study this potential effect, we assess the performance of \HISIGMA and the comparison strategies under data-MC mismodeling.
We mimic data-MC differences by slightly distorting the the background 'simulation' sample used to train the DNN classifier.
We then use these classifiers to perform inference on the original data, with no distortion.
For the binned DNN score fit, we still assume the shape of the background DNN score shape is known exactly from some data-driven method. 
Therefore, inference with this distortion-trained classifier is not biased, just sub-optimal. 
In contrast, \HISIGMA learns the background density directly from the data, and is therefore unaffected by such background `simulation' mismodelings. 

\begin{figure}
    \centering
    \includegraphics[width=0.49\linewidth]{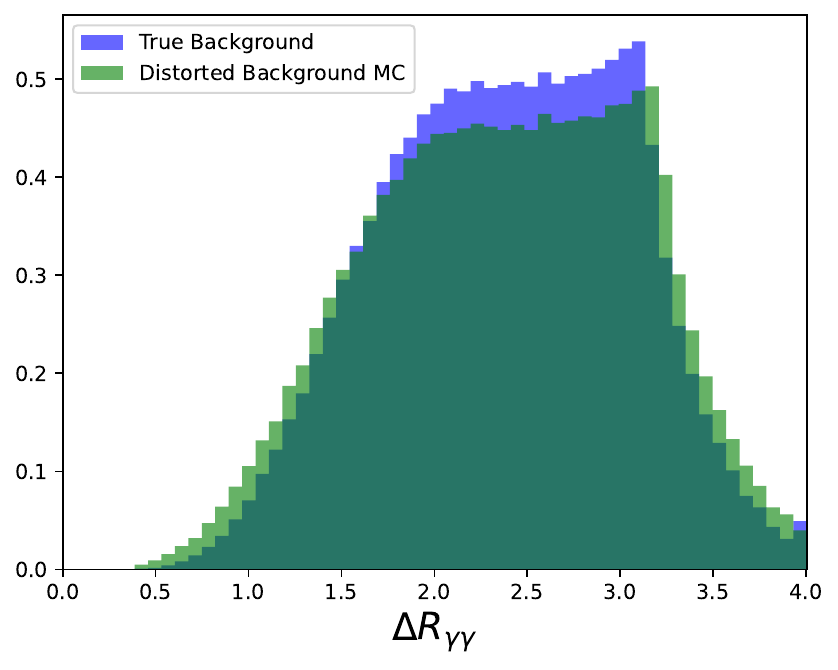}
    \includegraphics[width=0.49\linewidth]{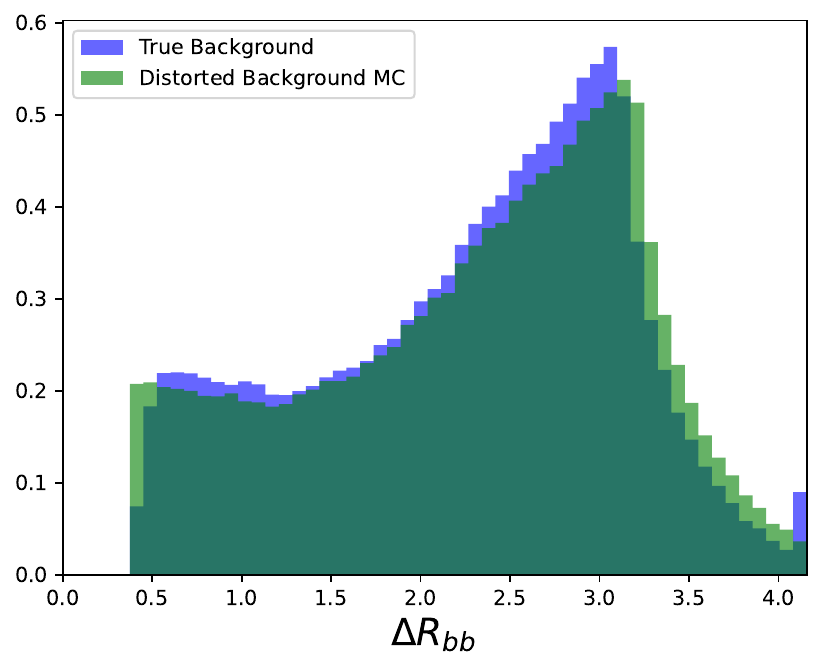}
    \includegraphics[width=0.49\linewidth]{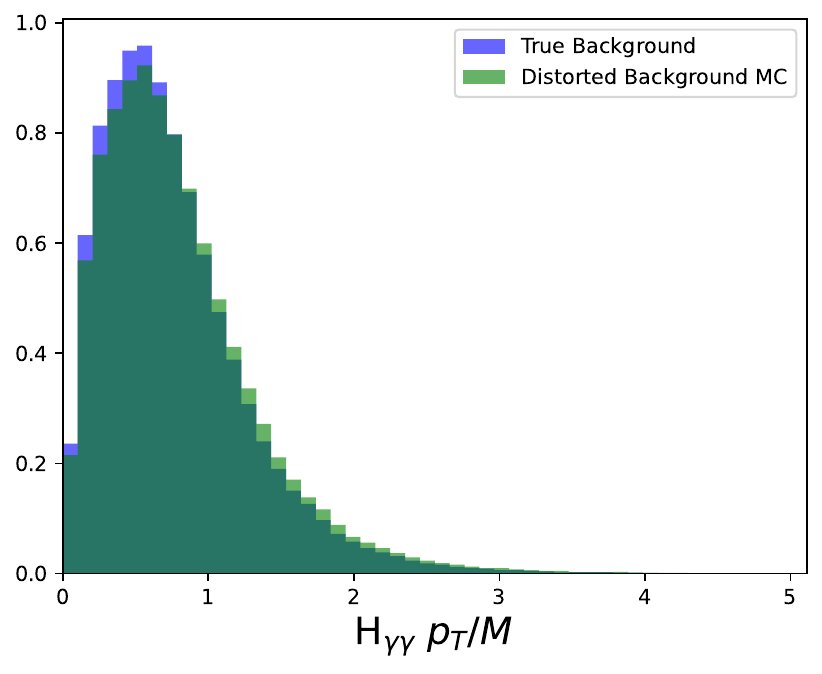}
    \includegraphics[width=0.49\linewidth]{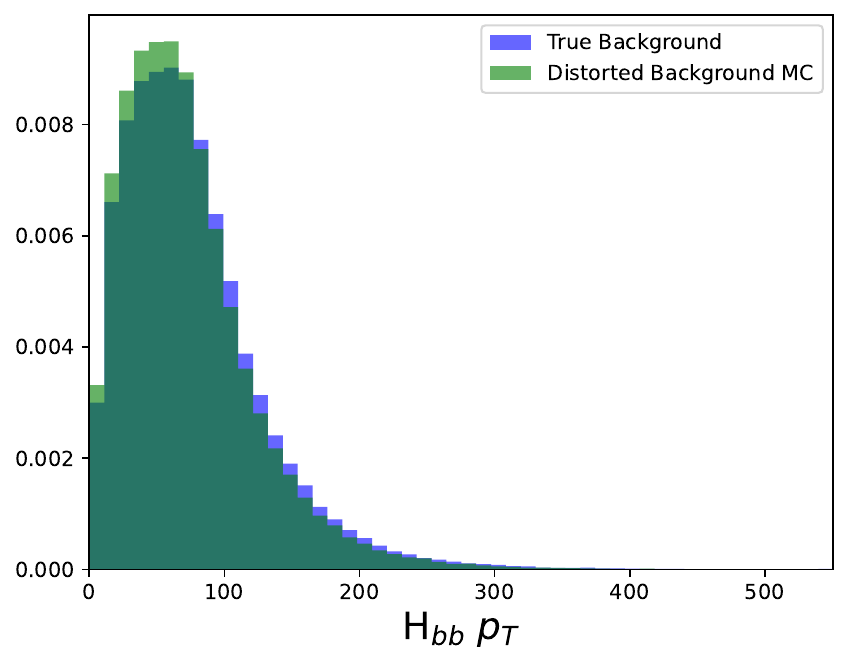}
    \caption{Illustrations of the original (blue) and distorted (green) feature distributions of the background sample. The distorted sample is used to train classifiers, approximating data-MC modeling imperfections. See text for details.}
    \label{fig:distortions}
\end{figure}

The distortions applied in our mock study, shown in Fig. \ref{fig:distortions}, are as follows.
The widths of the $\Delta R_{bb}$ and $\Delta R_{\gamma\gamma}$ distributions are increased by $\sim$10\%.
The H$_{bb}$ $p_T$ and H$_{\gamma \gamma}$ $p_T/m$ distributions have their means shifted by 5\% and are given 3\% additional Gaussian random noise.

\section{Results}
\label{sec:results}
\begin{figure}
    \centering
    \includegraphics[width=0.49\linewidth]{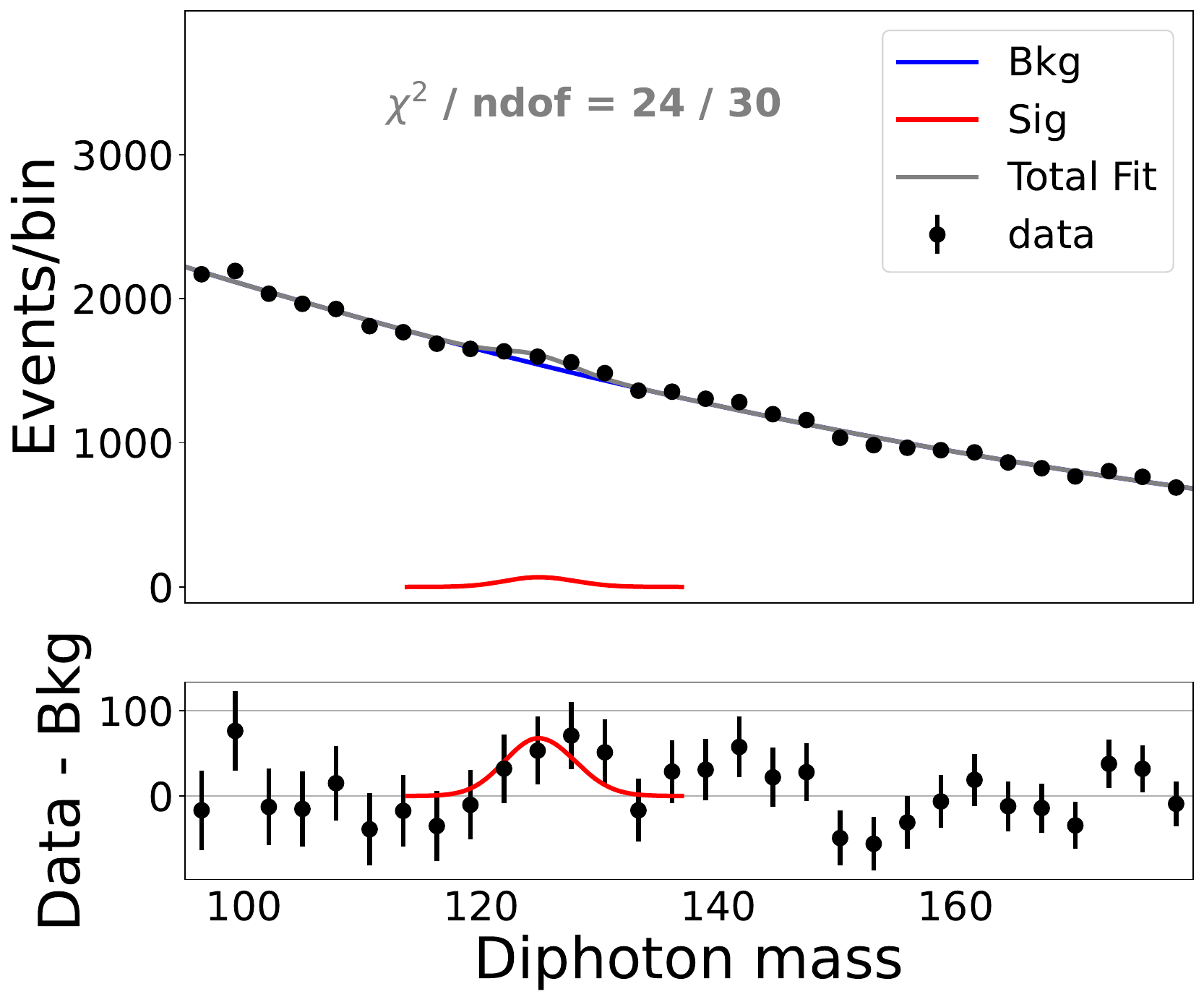}
    \includegraphics[width=0.49\linewidth]{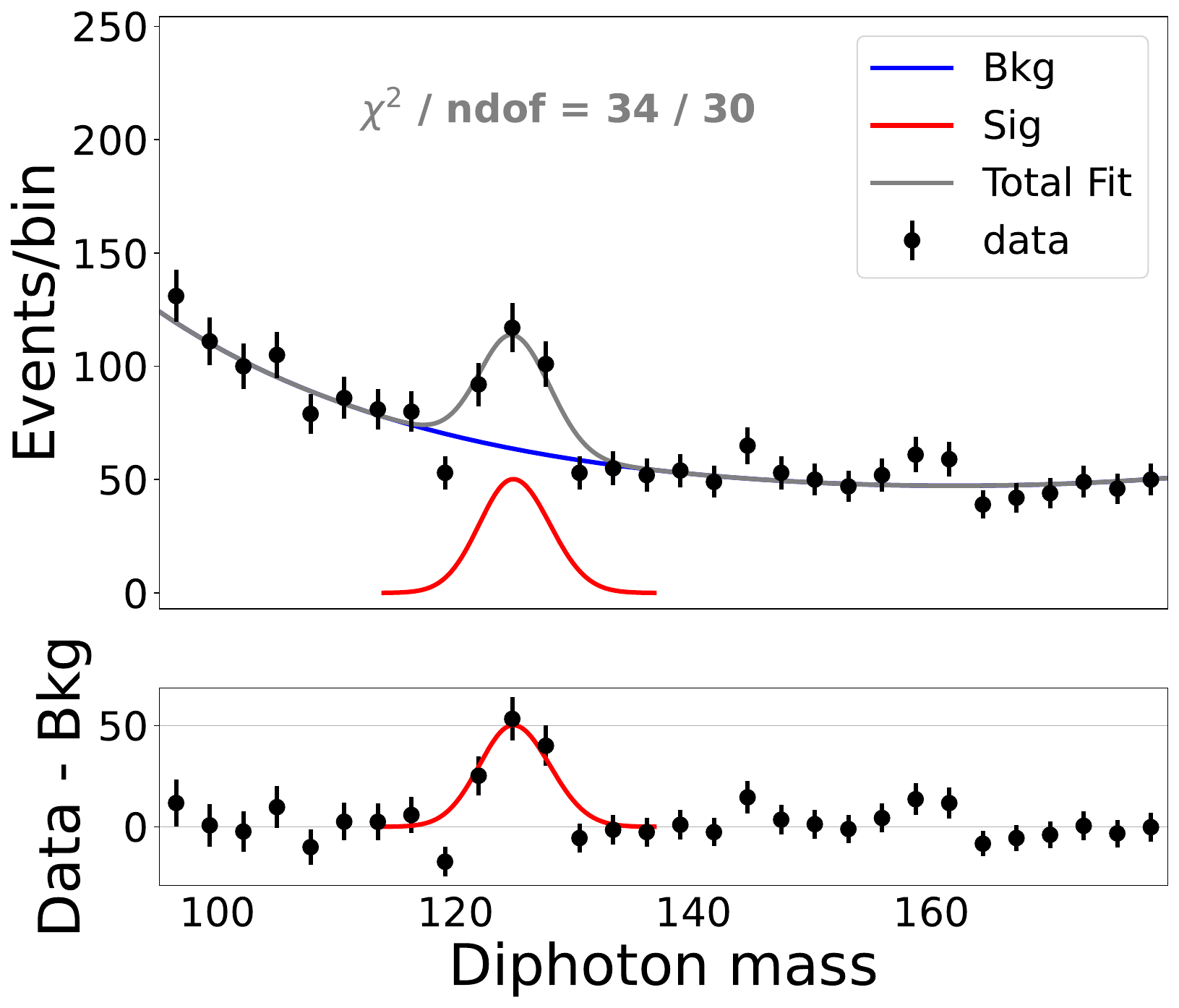} \\\
    \includegraphics[width=0.49\linewidth]{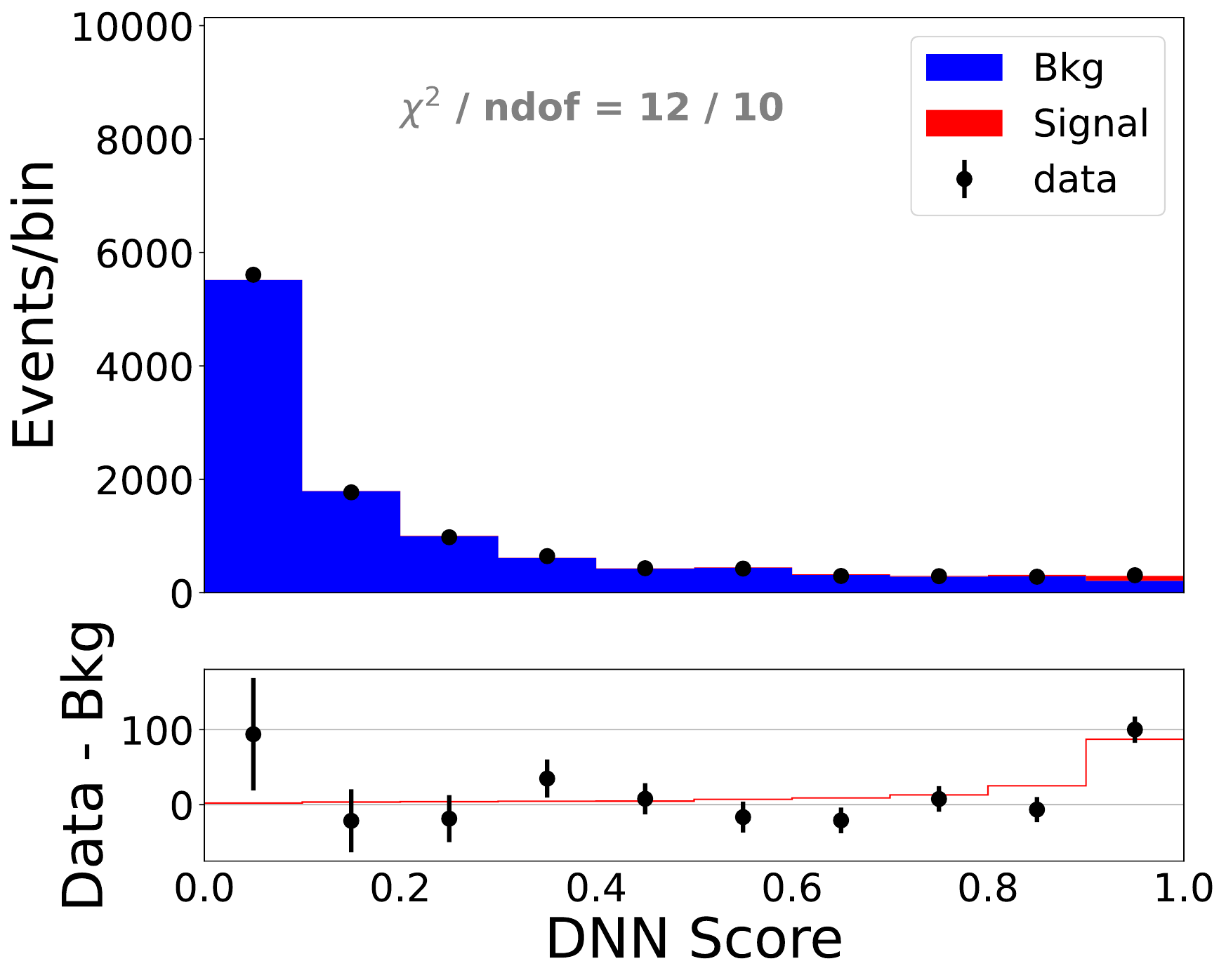}
    \includegraphics[width=0.49\linewidth]{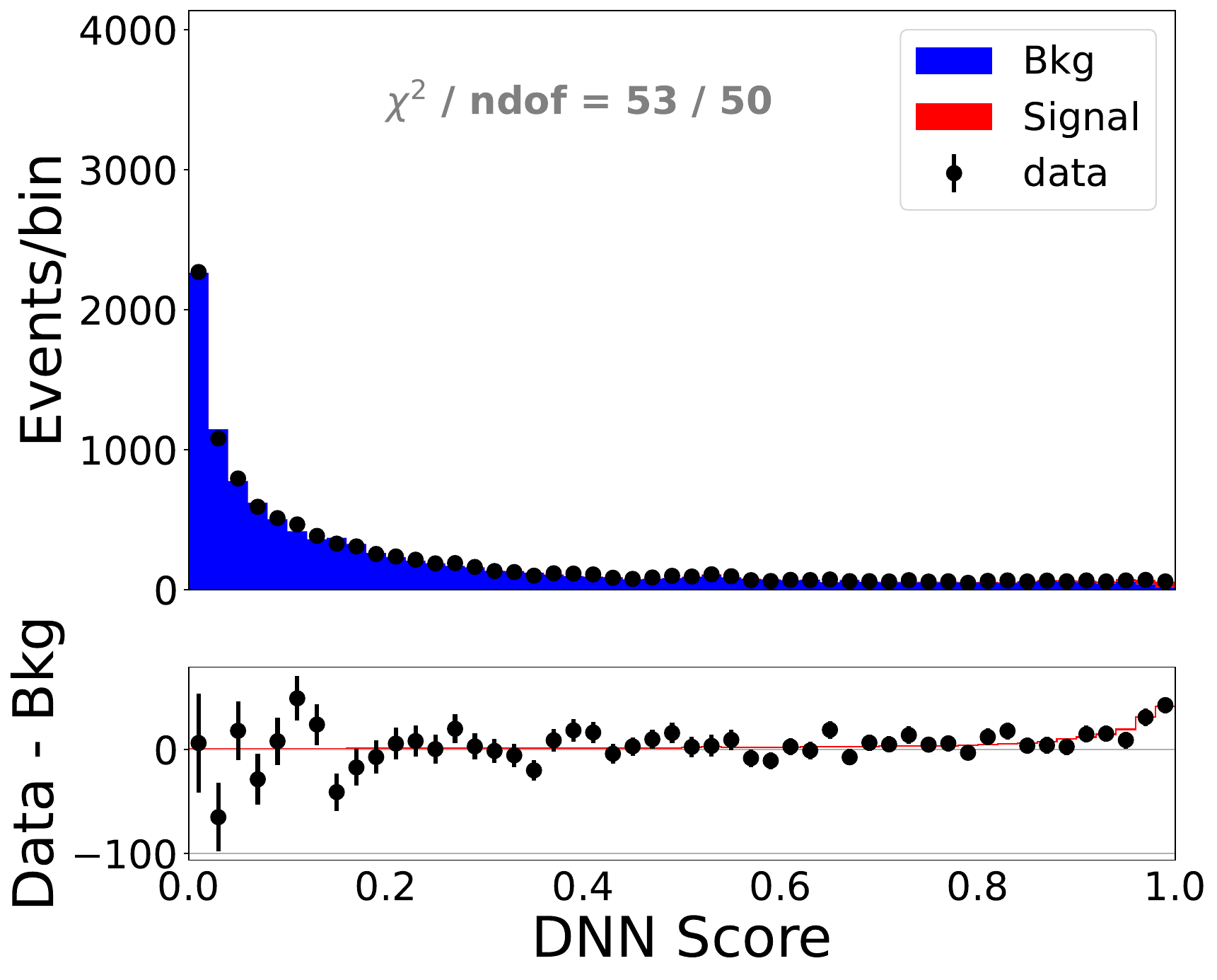}
    \caption{Examples of the mass only fit (upper left), cut+fit approach (upper left) and classifier score fit with 10 (lower left) and 50 (lower right) bins.}
    \label{fig:DNN_fits}
\end{figure}

The different analysis methods detailed in Section \ref{sec:dihiggs} are evaluated on 5 independent data samples, each containing 40,000 events.
Unless otherwise noted, for each data sample 200 signal events are injected, leading to an approximate significance with the mass-only fit of ${\sim} 3 \sigma$.
The different analysis methods are then applied to these datasets and compared.
Examples of the mass only fit, classifier cut and fit, and binned DNN score fit are shown in Fig. \ref{fig:DNN_fits}.

\subsection{Visualization of the \HISIGMA Fit}

\begin{figure}
    \centering
    \includegraphics[width=0.49\linewidth]{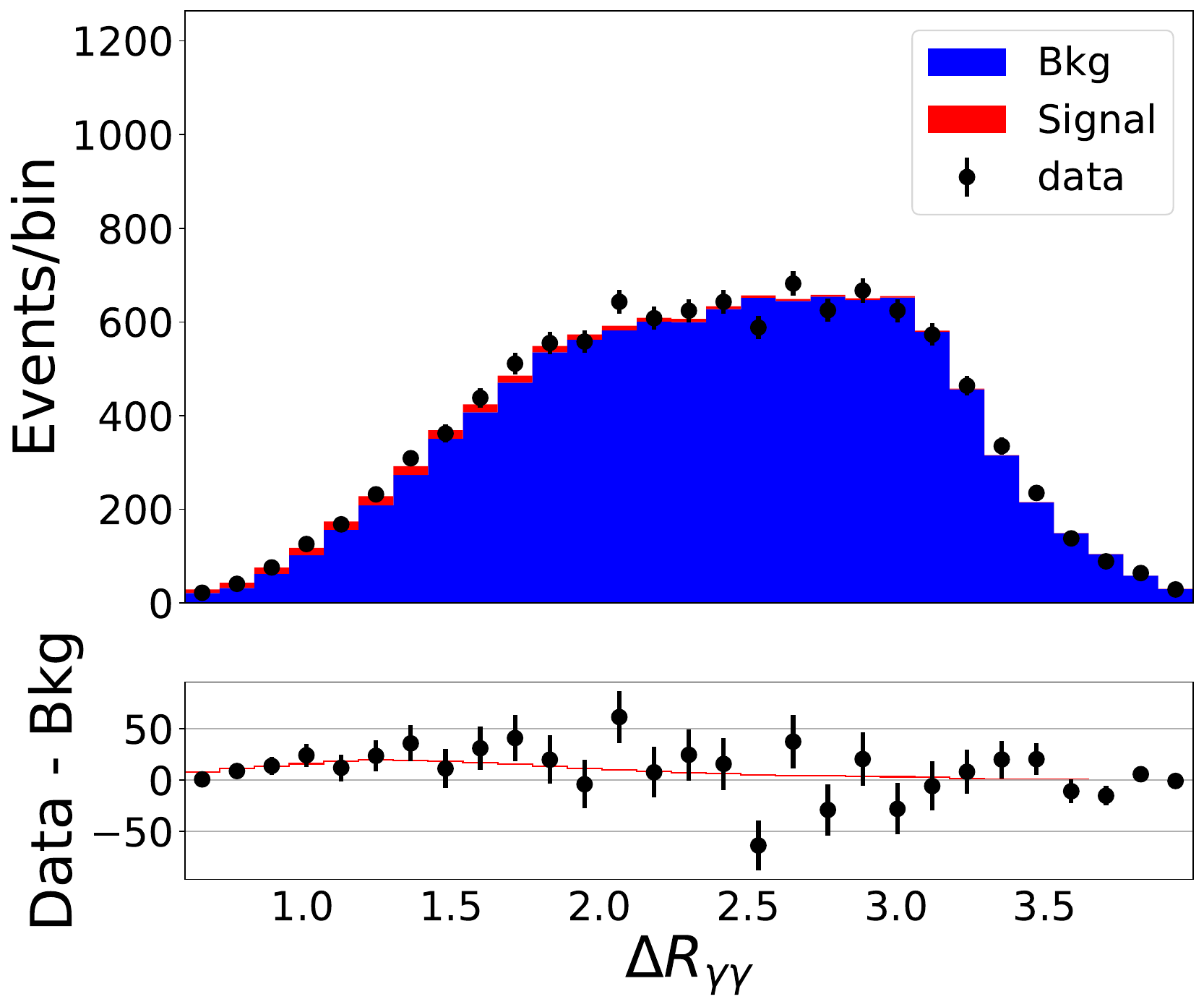}
    \includegraphics[width=0.49\linewidth]{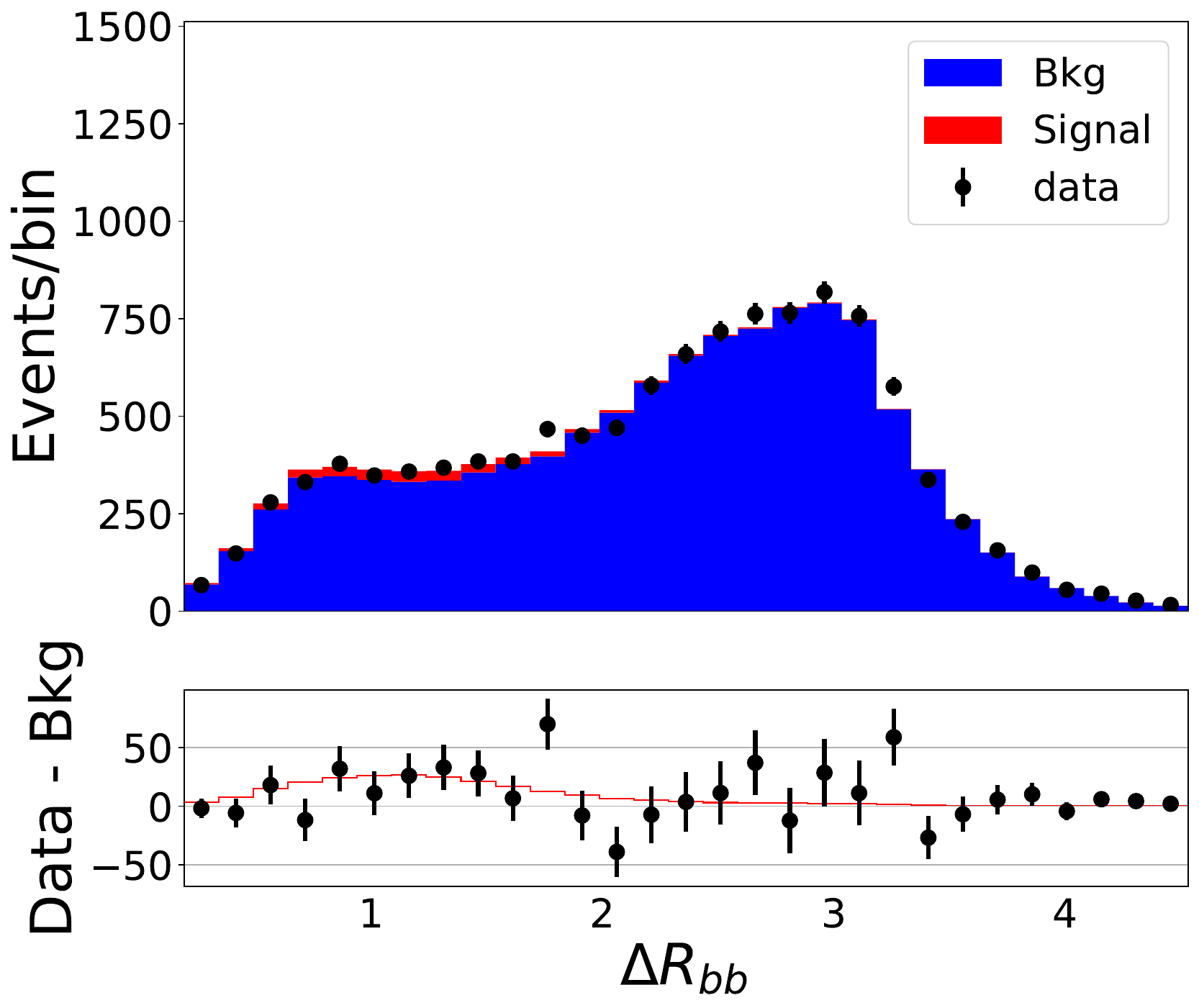}
    \includegraphics[width=0.49\linewidth]{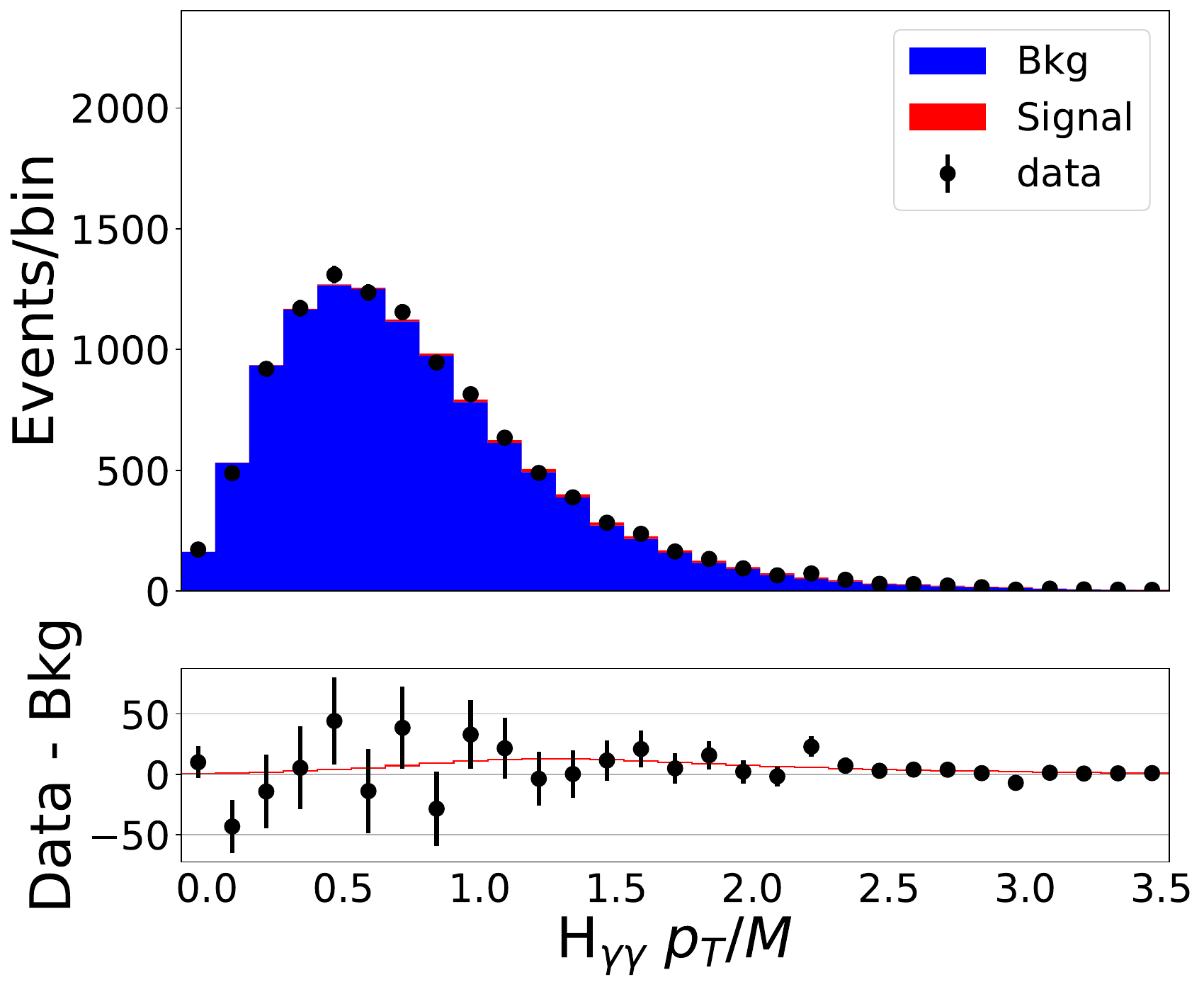}
    \includegraphics[width=0.49\linewidth]{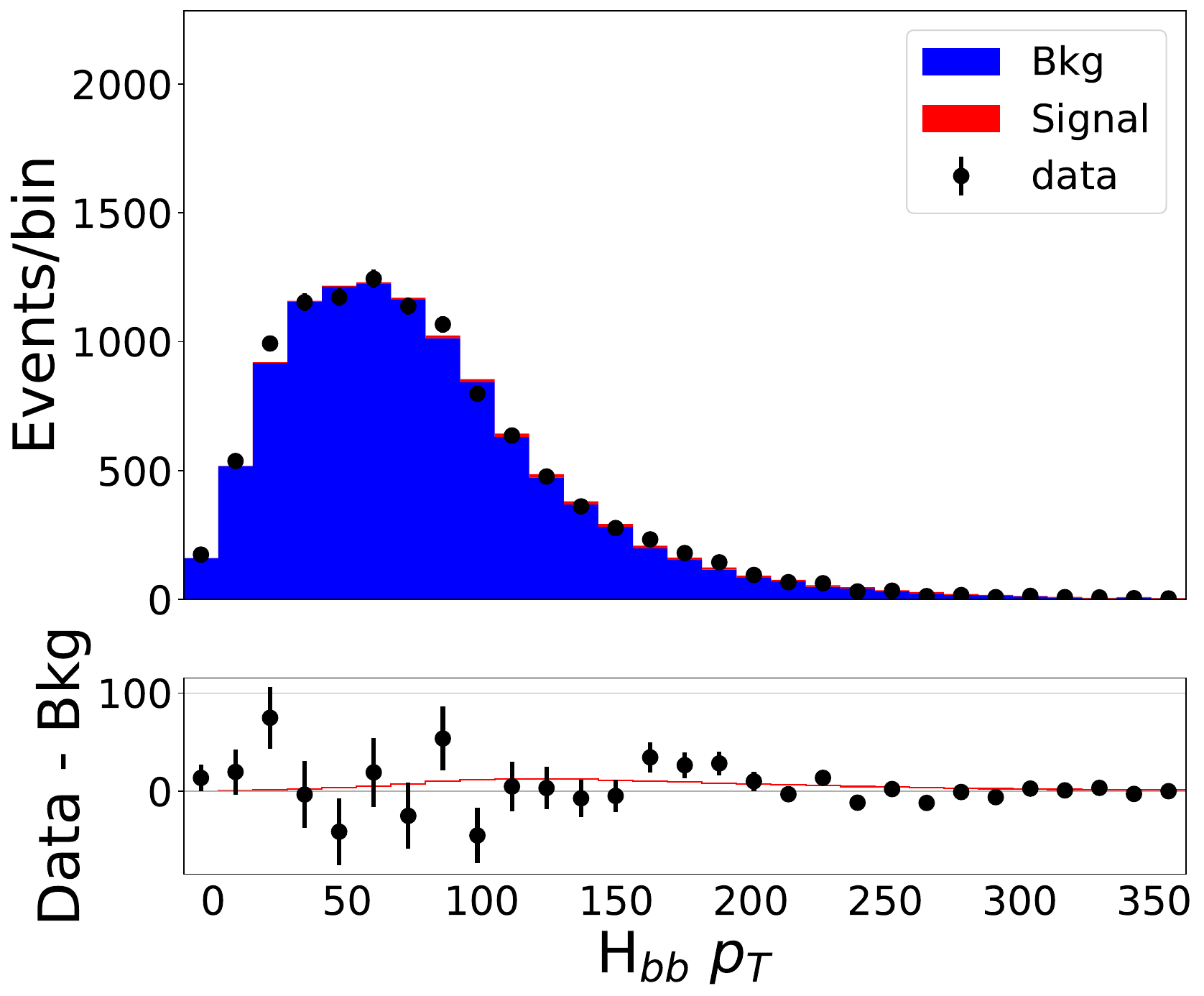}
    \caption{A visualization of the \HISIGMA fit across the various features. Though the fit itself is unbinned and multi-dimensional, binned 1D projections of the data and fitted \HISIGMA estimates are shown for visualization purposes.}
    \label{fig:HI_SIGMA_fits}
\end{figure}

The \HISIGMA fit across the multi-dimensional space can also be visualized. 
To do this, we sample events from the signal and background models to construct `templates' of our signal and background distributions. 
These templates are then scaled by the best fit normalizations and compared to the observed data events. 
This comparison can be done in any way that is desired.
The simplest choice is to consider 1D projections of each feature, which we visualize as histograms.
Examples are shown in Fig. \ref{fig:HI_SIGMA_fits}.
We stress that the binning is for visualization purposes only, as the actual model and fit are unbinned. 
For each feature one can see an excess of data events above the background model which is well-fit by the signal template.

\begin{figure}
    \centering
    \includegraphics[width=0.49\linewidth]{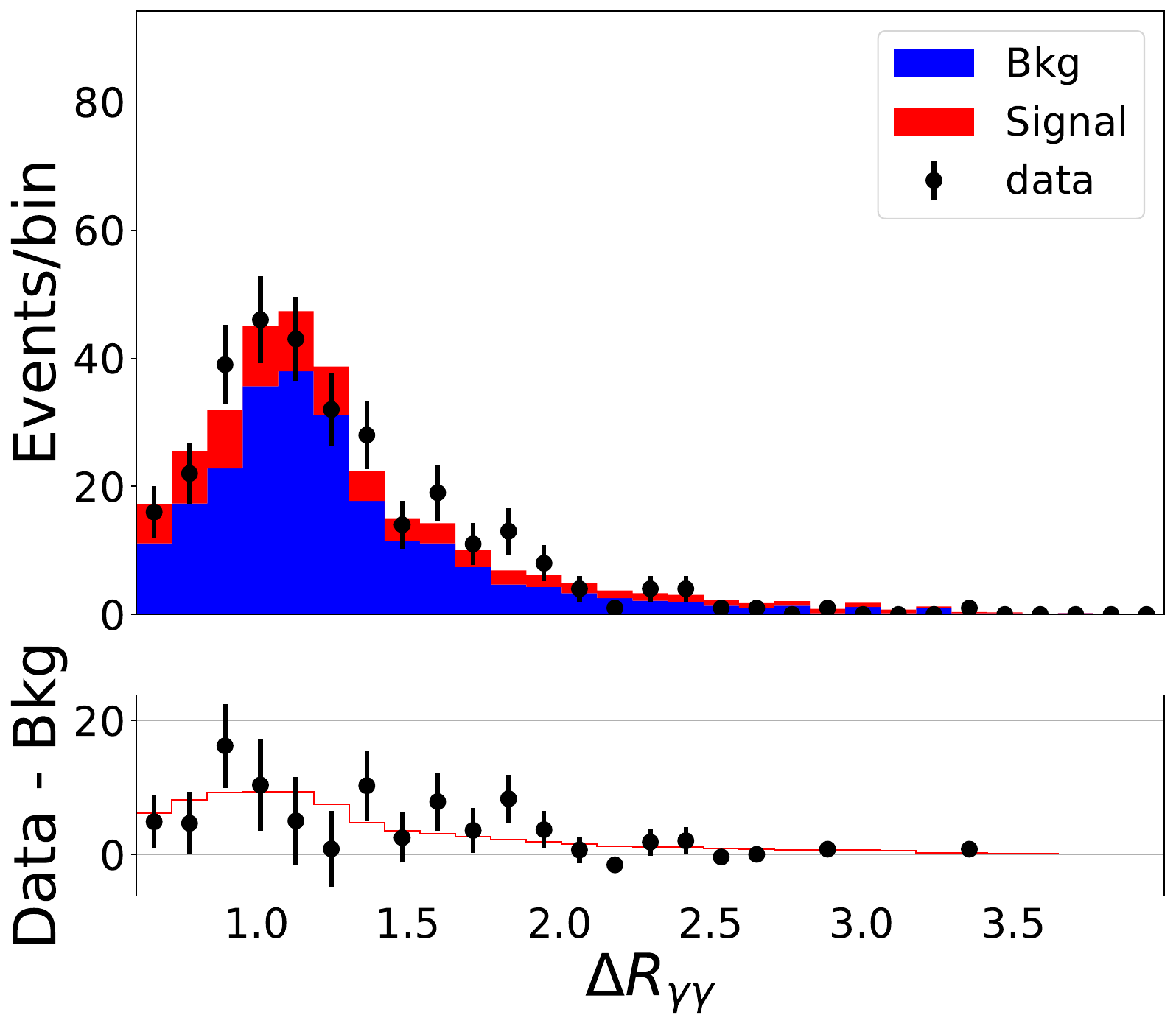}
    \includegraphics[width=0.49\linewidth]{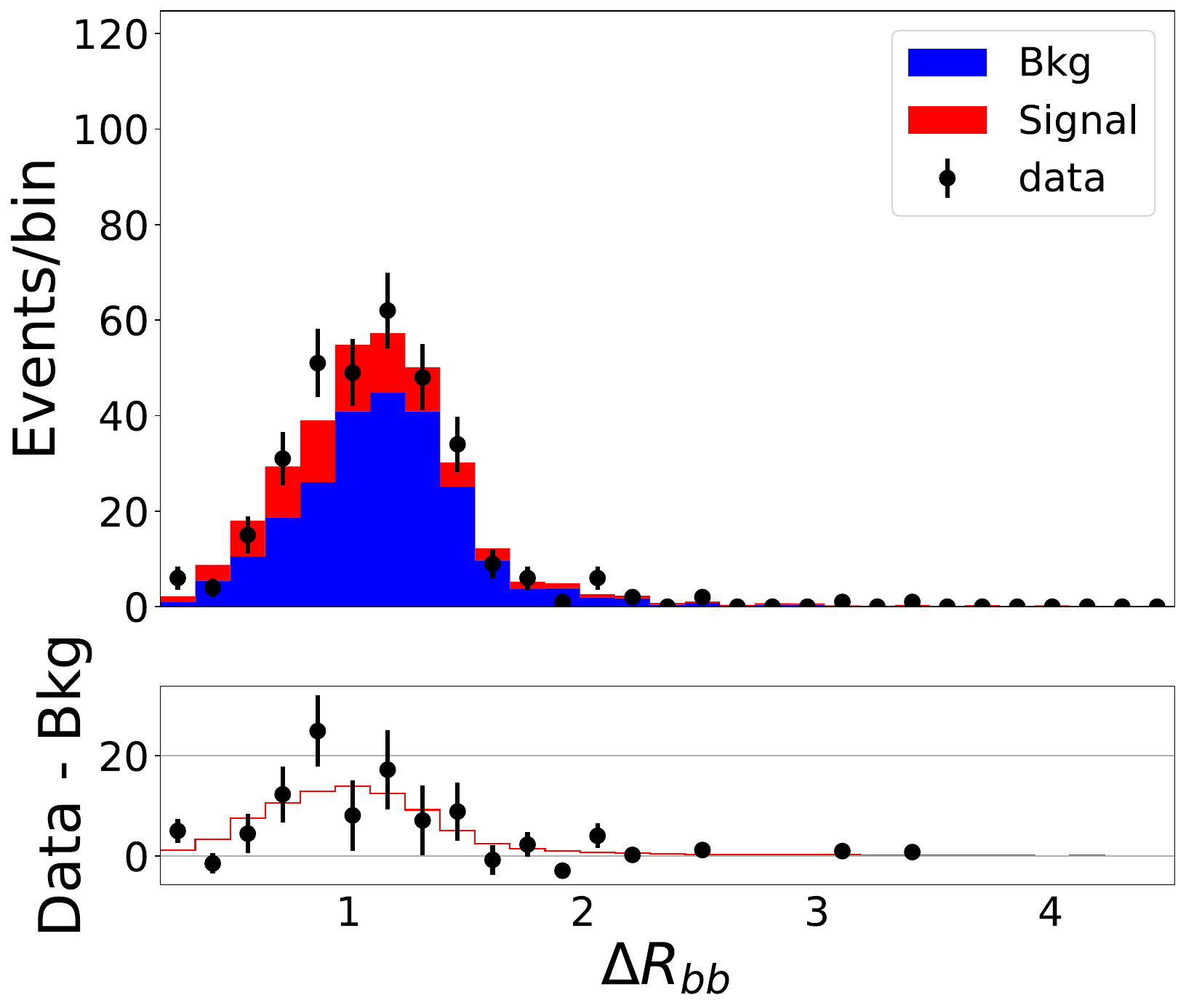}
    \includegraphics[width=0.49\linewidth]{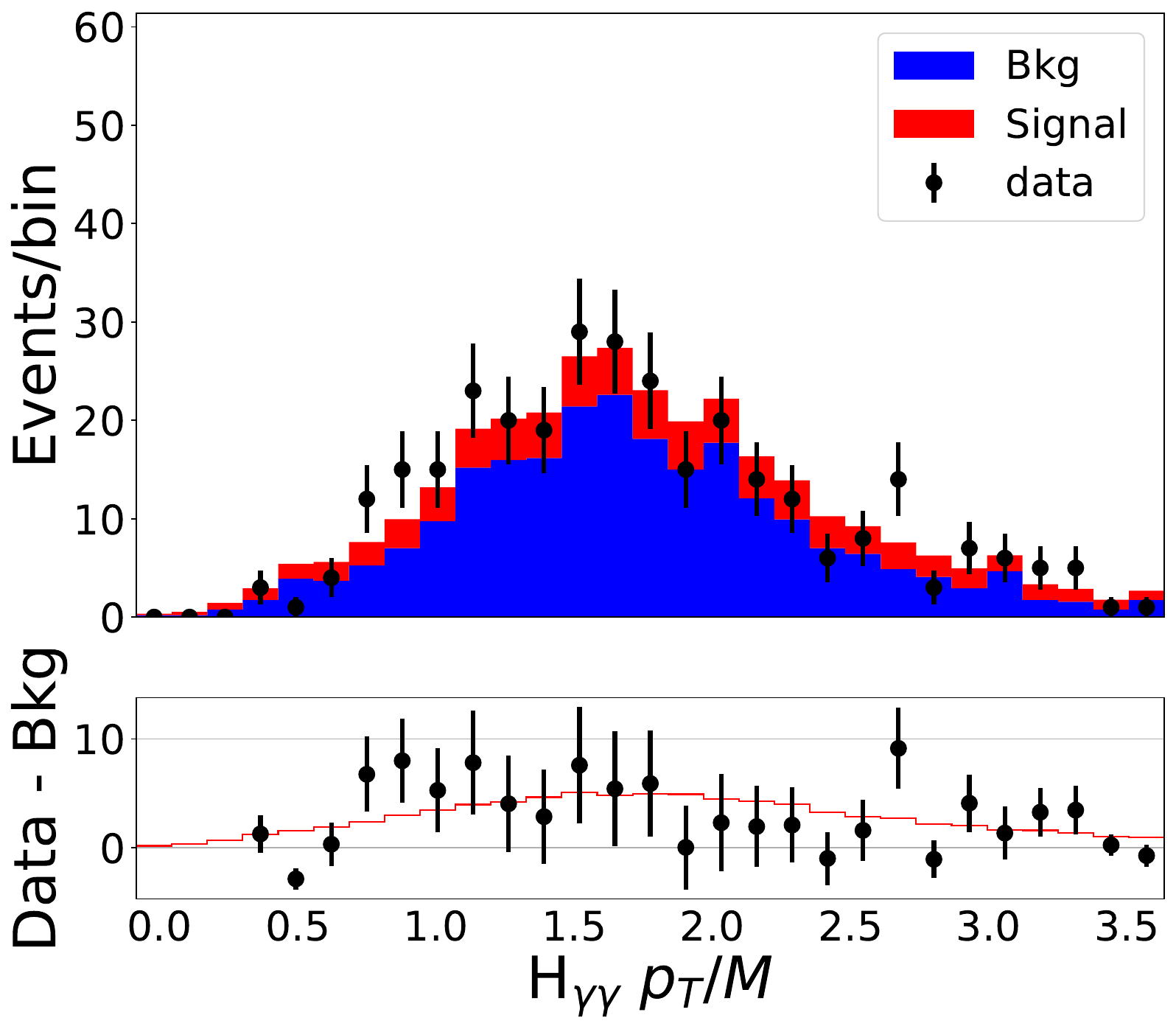}
    \includegraphics[width=0.49\linewidth]{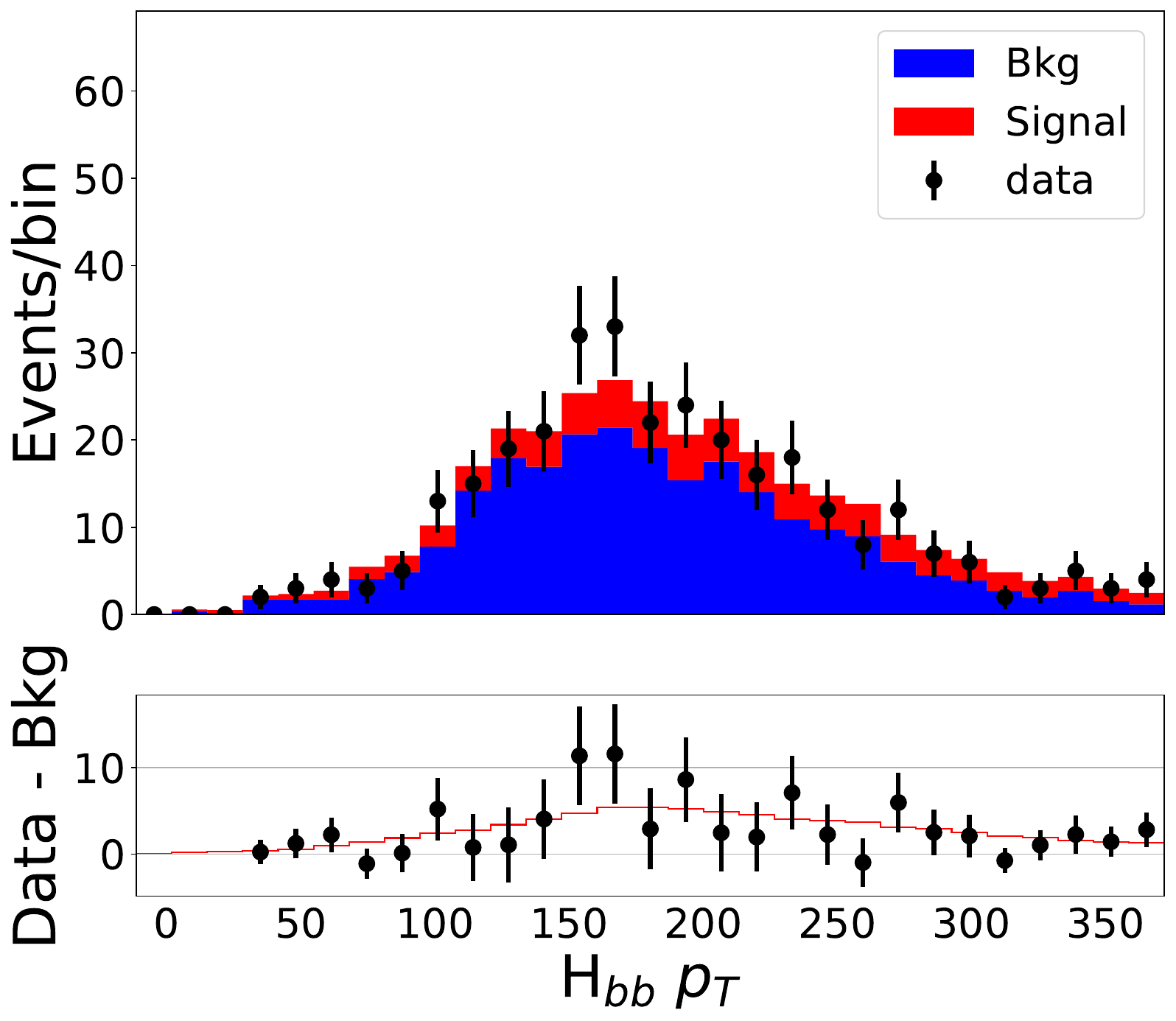}
    \caption{Visualization of a subspace of the \HISIGMA fit, the most signal-sensitive region, defined by a requirement on the supervised classifier score $> 0.9$. }
    \label{fig:HI_SIGMA_zoom_fits}
\end{figure}

While the visualization in Fig. \ref{fig:HI_SIGMA_fits} is useful, these 1D projections do not ensure the data is being well described across all regions of the multi-dimensional space. 
Fortunately, because \HISIGMA includes a full multi-dimensional model of the signal and background densities, one can check the fit agreement in sub-regions of the space easily. 
We consider a check of the most signal-sensitive region of the space, defined by placing a cut on the DNN classifier score $> 0.9$ and restricting $m_{\gamma\gamma}$ to the signal region, 115--135 GeV. 
Placing these cuts on the events generated from our model as well as the data allows us to check modeling in this sub-region of the space.
Such a comparison is shown in Fig. \ref{fig:HI_SIGMA_zoom_fits}. 
We see that in this sub-region the data is well described by the model and signal contribution is much more visible. 
We stress that such a visualization can be performed for any sub-region of the feature space $\vec{x}'$, but cannot be done for any region defined by an observable outside the feature space of the model (\textit{e.g.} checking modeling in different regions of $\eta$ for our example). 
Although this method is helpful in assessing the quality of the fit in different regions of the multi-dimensional space, a full assessment of the goodness-of-fit of an unbinned high-dimensional fit would require further study~\cite{Williams:2010vh}.

\subsection{Performance Quantification}

\begin{figure}
    \centering
    \includegraphics[width=0.6\linewidth]{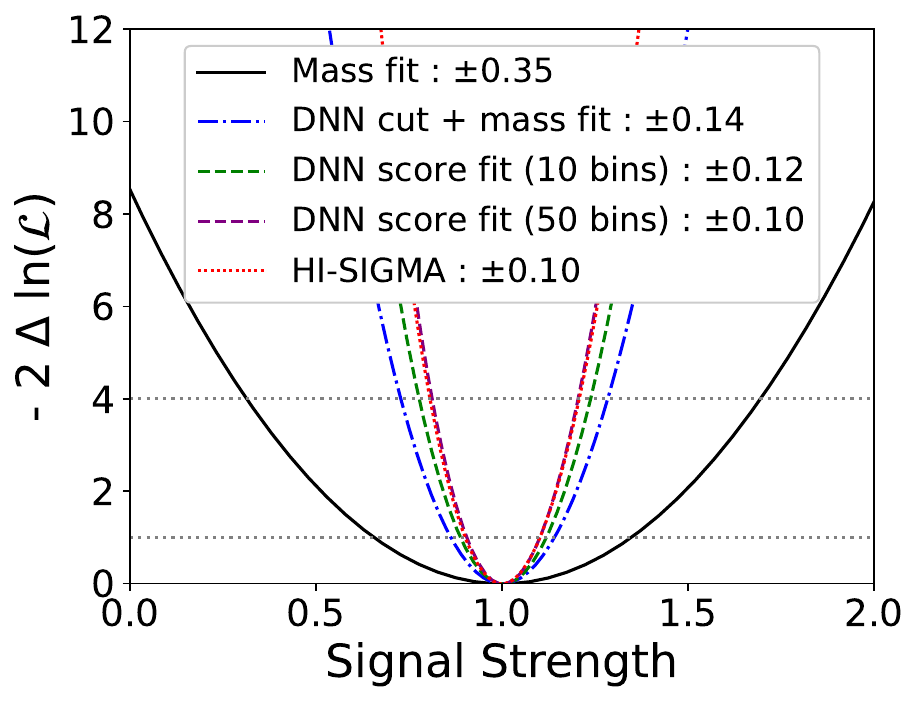}
    \caption{A comparison of the sensitivities of the different approaches. Averaged over 5 different datasets. Only statistical uncertainties are considered.}
    \label{fig:likelihood_cmp}
\end{figure}

The average uncertainty on the signal strength parameter (defined as $\frac{s}{s_{\mathrm{true}}}$) from the fit to the 5 independent datasets is summarized in Fig. \ref{fig:likelihood_cmp}. 
The \HISIGMA approach is shown to improve upon the sensitivity of the cut+fit approach by  $\sim 40\%$, and by $\sim 20\%$ as compared to the classifier score fit with 10 bins.
The performance of \HISIGMA is similar to the classifier score fit with 50 bins. 
We stress that this is an `idealized' comparison, in which no systematic uncertainties have been included for any of the analysis strategies, and strategies for realistic estimates of the background in the classifier score fit have not been considered.
The different analysis strategies would necessitate different systematic uncertainties, which may change their relative sensitivities. 
However, the ability of \HISIGMA to match an idealized DNN score fit with a large number of bins (50), while including a realistic background estimation strategy, demonstrates the potential of the approach.

\subsection{Robustness to Simulation Mismodeling}
\label{subsec:distortion}


Another advantage of \HISIGMA as compared to classifiers is that it learns the background density directly from data and thus avoids any loss of optimality due to imperfect MC background simulation. 
Since the classifiers are trained on MC, any data-MC difference will result in a suboptimal classifier score and a consequent degradation in sensitivity. 
This effect persists even if data-driven methods are employed to estimate the background for the classifier-based analyses.
To illustrate this, we perform a mock study where we construct a training sample for our classifier with slightly distorted features and then use the resulting trained classifier for inference on un-distorted event samples.

\begin{figure}
    \centering
    \includegraphics[width=0.6\linewidth]{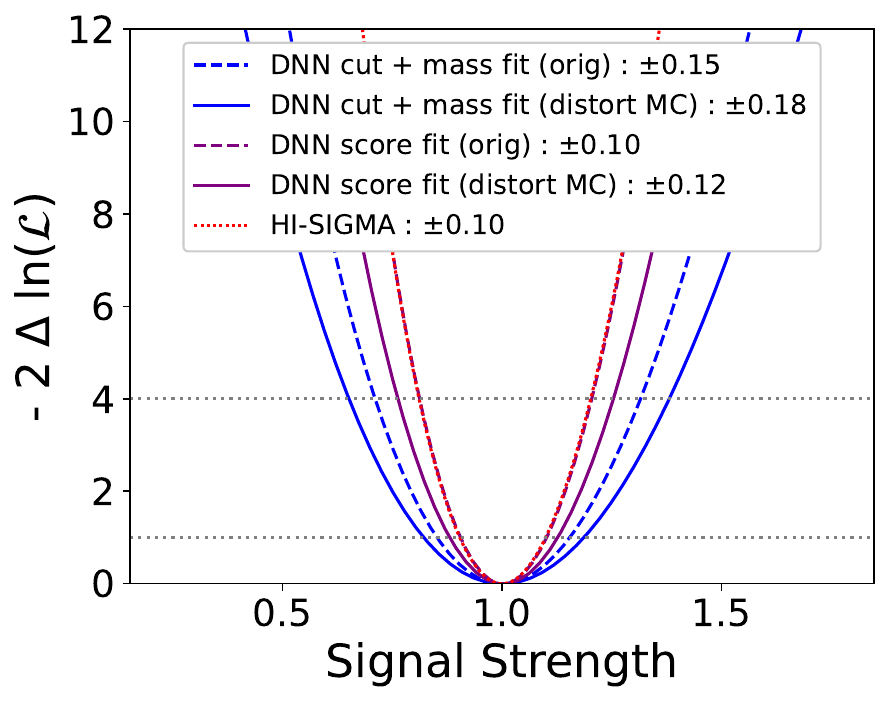}
    \caption{A comparison of the precision of the classifier-based approaches trained with distorted MC as compared to their original performance and \HISIGMA.
    The curve shown for the DNN score fit is based on the 50 bin version.
    The mismodelings of the background MC used in the classifier training reduces the sensitivity of the DNN-based approaches by ${\sim}$20\%, whereas \HISIGMA is unaffected.
    }
    \label{fig:likelihood_cmp_distort}
\end{figure}

Supervised classifiers are trained using these samples using the same architecture and hyperparameters discussed in Section~\ref{subsec:comparison_analysis_strategies}.
These classifiers are used to perform inference on the undistorted event samples. 
The results of these inferences are shown in Fig.~\ref{fig:likelihood_cmp_distort}. 
We observe that the training distortion degrades the sensitivity of the classifier-based methods by $\sim$20\%.
In contrast, as \HISIGMA does not rely on background simulation at all, it is not affected by these distortions.

For the cut+fit approach, because of our limited dataset size and reduced feature space, the optimal classifier selection ends up being quite loose, rejecting only 95\% of background events.
The efficiency of this loose selection is therefore relatively unaffected by the MC distortion, even though the classifier itself has reduced performance. 
For the comparison in Fig.~\ref{fig:likelihood_cmp_distort} we therefore employ a slightly tighter classifier cut, rejecting 99\% of background, to demonstrate the effect.
This tighter selection has slightly better $s/\sqrt{b}$ but slightly worse sensitivity ($\pm 0.15$) as compared to the optimal selection ($\pm 0.14$), due to larger uncertainties on the background shape arising from the reduced event count.
We note that current $bb\gamma\gamma$ analyses from ATLAS and CMS~\cite{ATLAS_bbgg, ATLAS_bbgg_run3, CMS_bbgg} cut very tightly on the classifier score, leaving only ${\sim}10$ events in the final signal region, and therefore could certainly be impacted by data-MC differences.

\subsection{Impact of Systematic Uncertainties}

\begin{figure}
    \centering
    \includegraphics[width=0.6\linewidth]{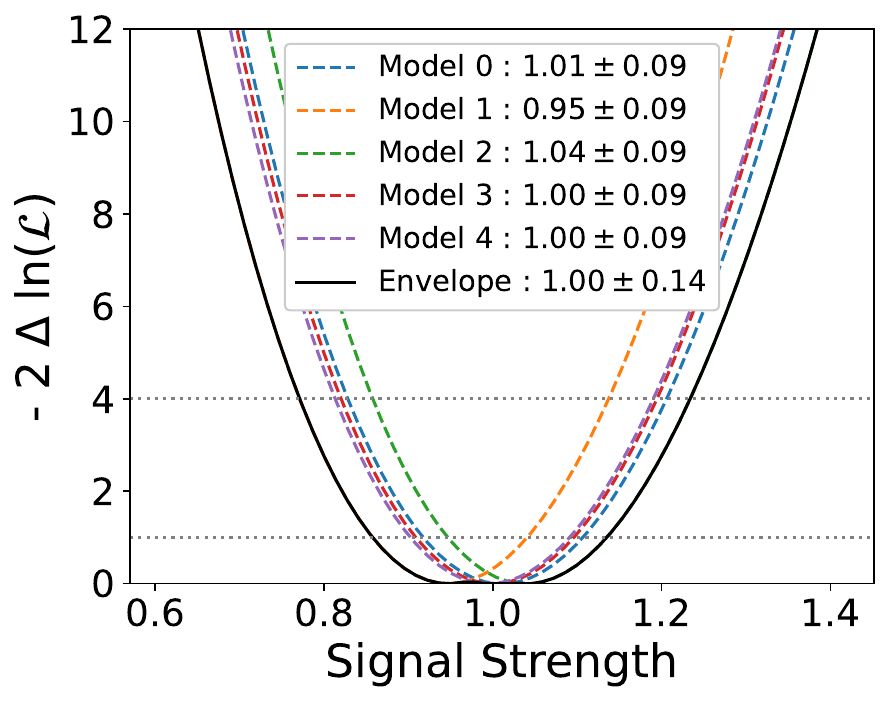}
    \caption{A demonstration of our `envelope of likelihoods' approach on a single mock dataset. The final \HISIGMA result is taken as the envelope of the different zero-aligned $\Delta \ln(\mathcal{L})$ of the five  models trained on different bootstrapped background samples.}
    \label{fig:profile_demo}
\end{figure}

We implement the likelihood envelope procedure described in Section \ref{subsec:finite_sys}, repeating the fit with the different bootstrapped models. 
The envelope is seen to increase the signal strength uncertainty by approximately ${\sim} 40\%$.
An example for the fit results on a single mock dataset is shown in Fig. \ref{fig:profile_demo}. 
It is unclear how much of the variation from the different bootstrapped models is coming from the finite statistics in the SB regions, and how much from intrinsic stochasticity of the generative model training process. 
If training hyperparameters were optimized, it is possible that stability could be improved, leading to a significant reduction in this uncertainty.

\begin{figure}
    \centering
    \includegraphics[width=0.6\linewidth]{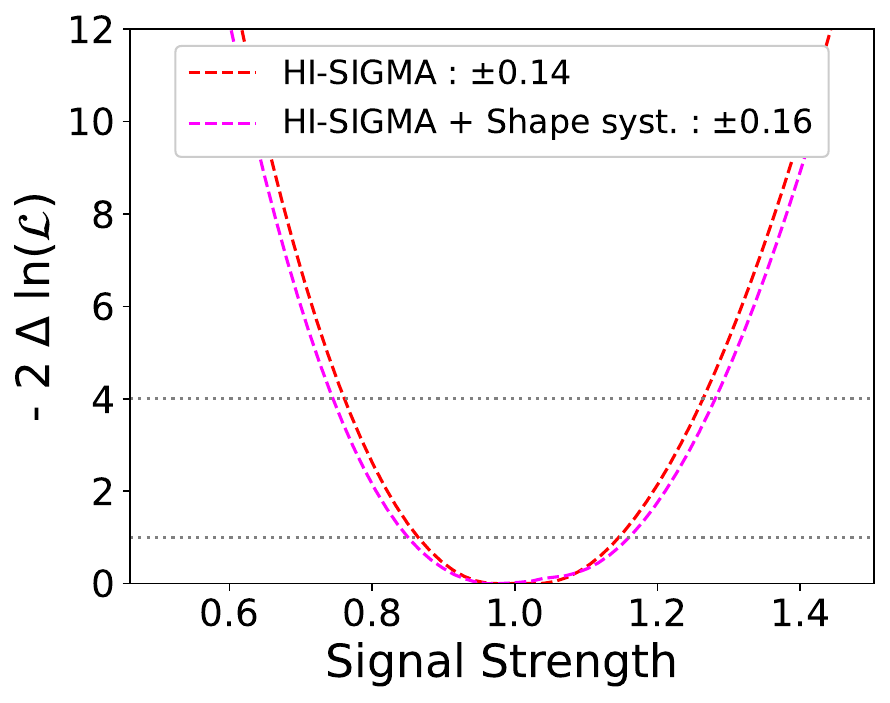}
    \caption{A demonstration of the impact of including the shape uncertainty on \HISIGMA. Both curves show the likelihood contour averaged over 5 different datasets and with `the envelope of likelihoods' method included.}
    \label{fig:likelihood_cmp_with_syst}
\end{figure}

We also implement the toy shape systematic described in Section \ref{subsec:shape_sys}.
The resulting impact from its inclusion is shown in Fig.~\ref{fig:likelihood_cmp_with_syst}. In practice, we observe that the introduction of the shape uncertainty is noticeable but minor and affects the location of the best fit signal strength more than the associated uncertainty. Thus, to fully showcase its impact we combine it with the model uncertainty captured by bootstrapping. We observe how the introduction of the shape systematics increases the total uncertainty as expected.

Although we have validated the incorporation of shape systematics in toy data and in this simplified set-up, a detailed study of the impact on the \textit{p.o.i.} fit and the pulls on the associated nuisance parameters when including multiple systematics requires a more involved training and deployment of models. Thus, we leave such a study for future work and focus here on introducing the method and showcasing its possibilities.

\subsection{Coverage Studies}
\label{subsec:coverage}

Full coverage studies with large samples of independent datasets would be necessary to assess whether the envelope uncertainty is appropriate, but we performed some basic checks based on the 5 mock datasets of size 40k. 

\begin{figure}
    \centering
    \includegraphics[width=0.5\linewidth]{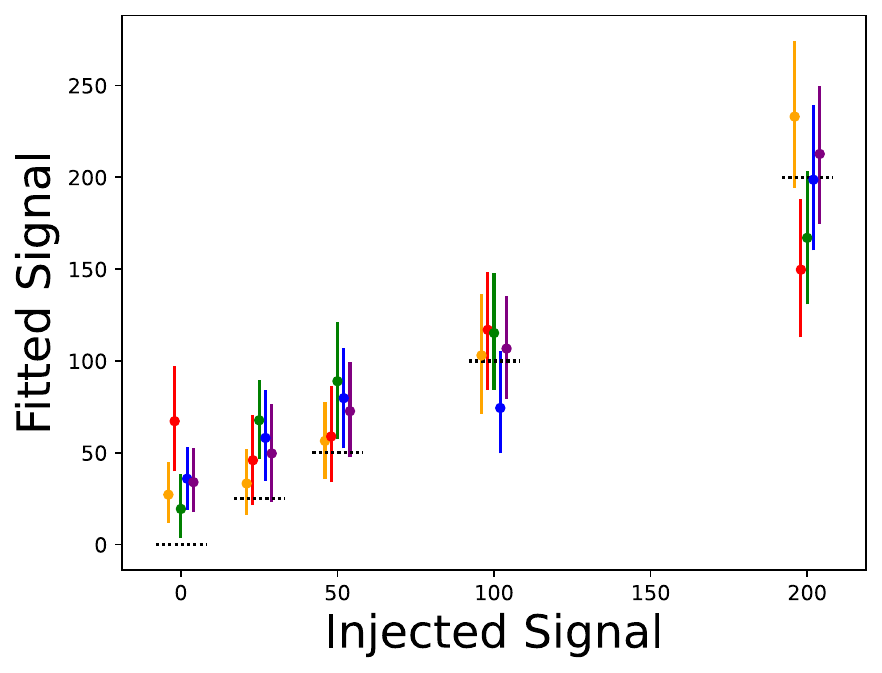}
    \caption{The extracted number of signal events for different injection tests of \HISIGMA on five independent datasets of size 40k. For larger signal injections, the measured values appear to adequately contain the true values, while for smaller signals a bias becomes more evident.}
    \label{fig:signal_size_scan}
\end{figure}

First, we consider five different signal strengths, varying from zero injected signal events up to the 200 events considered in the previous studies.
For each signal strength we inject a random selection of signal events into each mock dataset and perform inference. For simplicity, we include only the uncertainty associated with the finite training statistics, and do not consider the toy shape systematics. 
The results are shown in Figure~\ref{fig:signal_size_scan}.

We see that for the larger signal injections of 100 and 200 events, \HISIGMA seems to be performing well, with the true signal being within the uncertainties of most of the inferred signals and showing no obvious bias.
However at low signal injections a bias begins to appear, where the inferred signals are consistently higher than the true signal values. In particular, for zero injected signal the bias appears to be larger than $1 \sigma$.

It is likely that the bias arises from some mismodeling of the background by the normalizing flow model. If some region of phase space has an artificially high signal-to-background density ratio due to artificial background suppression, the fit would prefer to increase the fitted number of signal events in order to increase the likelihood of all background events that fall in that region. 
To check if this bias originates from issues in the interpolation of the background estimate from the sidebands into the mass signal region, separate background models were trained using events from the signal region. Inference performed with these background models were found to be similarly biased, indicating the bias does not appear related to the interpolation but instead seems to be a reflection of a limitation of the current density modeling. Further work would be needed to improve the background model and eliminate this bias.

To further test the coverage properties of \HISIGMA at low signal strengths we perform a more thorough coverage test using small event samples. 
We split our available testing set of 200k events into 50 independent samples of 4k events. 
For each sample we inject a known number of signal events, perform inference with \HISIGMA, and compare the measured signal strength to the true value.
We repeat this procedure for two signal strengths: a background-only test, where no signal events are injected, and for an injection of 25 signal events. 
The latter injection corresponds to approximately $2.5 \sigma$ of evidence, due to the reduced background statistics. For each measurement we compute the pull, defined as

\begin{equation}
    t = \frac{s_\mathrm{meas} - s_\mathrm{true}}{\sigma},
\end{equation}
for measured signal events $s_\mathrm{meas}$, true signal events $s_\mathrm{true}$ and measurement uncertainty $\sigma$. If \HISIGMA yields an unbiased and properly calibrated estimator, the pull distribution will be a Gaussian with mean zero and standard deviation of one. 
The 1-$\sigma$ coverage fraction can be assessed as the fraction of measurements with $|t| < 1$. A well-calibrated pull should yield a 1-$\sigma$ coverage fraction consistent with 68.3\%. 

In Table~\ref{tab:coverage} we report these quantitative metrics for \HISIGMA both with and without the finite training uncertainty described in Section \ref{subsec:finite_sys}.
We observe that for all four cases the pull distribution indicates no significant bias. 
Proper coverage is achieved when including the finite training uncertainty. 

The fact that no significant bias for the no signal case is observed on this smaller sample,
in contrast to the apparent bias seen on the larger sample, further reinforces the hypothesis that the source of previously observed bias resides in the background mismodelling.
The effect of this mismodeling becomes subdominant with respect to the increased statistical uncertainties of the smaller sample, ameliorating the bias.
This is encouraging, as it implies that the bias in the background estimation is likely small, and that \HISIGMA can perform reliable inference for moderate sample sizes of $\mathcal{O}(5\times10^{3})$, which are realistic for many analyses. 

We expect the particular range of trustworthy sample sizes to depend on the amount of training data and quality of training. In general, further improvements to the generative model used for the background estimate or better uncertainty parameterizations would allow unbiased inference on larger samples.
For a realistic applications, such as a $bb\gamma\gamma$ analysis, a classifier selection could be applied prior to \HISIGMA, and the threshold of the classifier chosen to result in a number of events amenable to accurate inference from \HISIGMA. 
If the classifier selection does not sculpt the mass distribution or introduce sharp features that are difficult to capture in the background model, \HISIGMA could be used in this signal-enriched dataset without any modifications.

\begin{table}[]
    \centering
    \begin{tabular}{c | c | c c c}
        Method & Injection size & Mean pull & Std. dev. pull & 1-$\sigma$ coverage frac. \\
        \hline
         \HISIGMA (no sys) & 0 & $-0.12 \pm 0.25$ & $1.72 \pm 0.17$ & $38\pm 9$\% \\
        & 25 & $-0.24 \pm 0.15$ & $1.03 \pm 0.10$ & $64\pm11$\% \\
        \hline
        \HISIGMA (w/ sys) & 0 & $0.14 \pm 0.15$ & $1.05 \pm 0.11$ & $64\pm 11$\% \\
        & 25 & $-0.19 \pm 0.12$ & $0.87 \pm 0.09$ & $72\pm 12$\% \\
         
    \end{tabular}
    \caption{Coverage test performed on 50 independent datasets of 4k events each. The pull distribution mean, standard deviation and 1-$\sigma$ coverage fraction are reported for two different signal injections. 
    The evaluations are performed both and without the systematic uncertainty arising from finite training statistics shown in Sec.~\ref{subsec:finite_sys}. Uncertainties on the reported values originate from the limited number of toys.
    Good coverage properties are seen for both injection sizes with the inclusion of systematic uncertainties.}
    \label{tab:coverage}
\end{table}

\section{Discussion}
\label{sec:discussion}

In this work we have introduced \HISIGMA as a new  multivariate analysis strategy and demonstrated its potential. There are several factors to consider when weighing the pros and cons of the \HISIGMA approach as compared to other analyses strategies. 
We offer here some discussion comparing \HISIGMA to the cut+mass and the binned classifier score fits. 

\subsection{Optimality}
As demonstrated in Section \ref{sec:results}, if using the same input variables, the \HISIGMA approach offers a significant sensitivity gain as compared to the cut and fit approach and a moderate sensitivity gain as compared to fits to a classifier score fit with a low number of bins.
Additionally, because the classifier based approaches are typically trained fully on MC, data-MC differences could cause degradation of the sensitivity.
In the \HISIGMA approach, the background probability density is learned from the data itself, making it unreliant on data-MC mismodelings of the background.
This was demonstrated in Section \ref{subsec:distortion}.

Furthermore, in the \HISIGMA approach, all nuisances are profiled over within the full multi-dimensional space, allowing them to be optimally constrained by the data. 
This makes the \HISIGMA approach maximally `systematics-aware', as compared to classifiers which generally lose their optimality in the presence of systematic uncertainties and must be trained in specific ways to mitigate this effect \cite{ATLAS_SBI_methods,CMS:2025cwy}. 

The biggest drawback of \HISIGMA as compared to classifier-based approaches may be the scaling to very high dimensional feature spaces. 
We have demonstrated \HISIGMA works well on 4+1 dimensional feature+mass space, however  classifier-based analyses often employ 20 or more variables to achieve maximum separation power.
Further studies would be required to see how high-dimensional the \HISIGMA approach could be pushed, but is reasonable to expect it would lose some performance as compared to classifiers for very high dimensional spaces. 
If this is the case, a hybrid strategy may be possible, in which a classifier is applied in a preselection using some portion of a large feature set, and then \HISIGMA is performed on the remaining events using only the highest-importance features. 

\subsection{Data-driven Background Estimation}
The \HISIGMA approach naturally incorporates a data-driven background estimate, and our initial studies demonstrate how its uncertainties can be parameterized.  
In the cut+fit approach, the background estimate is typically straightforward, just done through a parametric mass fit.
However caution must be taken to ensure the classifier score cut does not sculpt the mass distribution and cause a bias. 
This can be difficult to verify for very tight classifier score cuts, often used in high-purity signal regions, in which there are few remaining events. 
In the binned classifier score approach, the background estimate is less straightforward. 
Generally some analysis-specific methodology must be employed and a reliable strategy may not be possible in all cases.
Within a resonance search of the type considered in the \HISIGMA, some sort of interpolation of the background distribution from the sideband regions is needed, which may only be possible for specific types of generative models. 
The current $bb\gamma\gamma$ analyses from ATLAS and CMS \cite{ATLAS_bbgg, ATLAS_bbgg_run3, CMS_bbgg} use a cut+fit strategy, despite the known sensitivity increase from binned classifier score fits, likely because of these difficulties. 

\subsection{Interpretability}
The cut+fit approach is the most naturally interpretable, as all the information of the fit is contained in a single physical feature, and the community has a long history of interpreting and scrutinizing resonance plots. 
Some interpretability may be lost if there are many categories and/or very tight cuts that make an assessment of the background modeling difficult. However, these multiple categories can generally be combined into a single plot for visualization with $S/(S+B)$ weighting. 
Fits to classifier scores, on the other hand, are generally less interpretable: as the observable being fit has no physical basis, it is difficult to assess whether the signal and background match expectations and spot potential mismodelings. 
As we have shown the \HISIGMA approach is naturally interpretable, as each observable being fit is physical, and one can easily inspect sub-regions of the multi-dimensional probability distribution to check the agreement with data. 

Though they have generally not generally been reported, in principle, if one has a true multi-dimensional model of the background being used in a classifier score fit, similar visualizations could be possible.
Data-driven high-dimensional background estimates have been recently employed in some analyses, often by using a multivariate reweighting from a control region to the signal region using an ABCD-like setup \cite{ATLAS_bbbb, CMS_bbbb, CMS_bbbb_boosted}.
Visualization can then be performed by checking the agreement between the data and the signal plus background estimate in all classification features for events in the most sensitive classifier score bins. 
However, some data-driven classifier score analyses only estimate the binned classifier score distribution being fit, or perhaps only account for some of the feature correlations in the reweighting from a control region, in which case such visualizations may not be possible.

\subsection{Comparison to SBI}
As SBI and \HISIGMA both perform high-dimensional unbinned statistical inference, both theoretically offer optimal single and multi-parameter extraction.
The largest advantage of \HISIGMA as compared to SBI methods is the enabling of data-driven background modeling, which has not been considered in SBI approaches. 
This should enable the application of \HISIGMA to the numerous resonance searches and measurements requiring data-driven background modeling.

The largest advantage for the classifier-based approach of SBI may be practicality.
It is common wisdom that classifiers are easier to train than generative models. 
One explanation is that training a classifier on $N$ input features encodes a function $\mathbb{R}^N\to \mathbb{R}$, whereas typical generative models (such as normalizing flows or diffusion models) learn a more complex  $\mathbb{R}^N\to \mathbb{R}^N$ function. 
It is therefore unsurprising that classifiers take less computational resources to train, can be trained with fewer examples, and can more easily scale to large $N$. 
Further research is required to assess how large a feature set can be accurately modeled in the \HISIGMA approach, and assess how many features are required for optimal inference in many analyses. 
However, given previous applications of generative models, its likely that \HISIGMA could scale to $O(10)$ dimensional distributions which is sufficient for many analyses.

Within the SBI terminology, \HISIGMA would be considered a form of `direct likelihood estimation'~\cite{Papamakarios:2019ccu, Sluijter:2025isc}, however because the background likelihood is learned from data, the SBI moniker does not seem appropriate. 
The other used moniker, `likelihood-free inference' also does not seem appropriate as the factorization of the resonant component in \HISIGMA explicitly specifies the structure of the likelihood. 
So called `direct likelihood estimation' is also the \textit{modus operandi} of particle physics analysis, as all analyses using histograms or parametric functions to model signal and background probability densities fall into this category. 
It thus makes more sense to categorize \HISIGMA as a `high-dimensional statistical inference method', rather than a form of SBI.

By focusing on density estimation, \HISIGMA shares a commonality with standard analysis techniques (like histograms), that likelihood-ratio-based SBI does not. 
This may have sociological benefits.
Demonstrating reliable coverage of statistical inference performed with \HISIGMA  would be equivalent to demonstrating the generative model well describes the data within its uncertainties. 
Furthermore, methods to validate data-driven background estimates are widely employed in standard analyses and are therefore well understood by the community, which may ease adoption of \HISIGMA in experimental collaborations. 

\section{Conclusions and Outlook}
\label{sec:conclusions}

In this paper we have introduced \HISIGMA, a new analysis methodology tailored to resonances, enabling unbinned high dimensional statistical inference with data-driven background modeling. 
Using as an example di-Higgs measurement in the $bb\gamma \gamma$ final state, we have demonstrated that \HISIGMA provides sensitivity gains over commonly employed classifier-based analysis strategies, while providing a data-driven background estimate and maintaining interpretability. 
Because \HISIGMA performs statistical inference in a high-dimensional space, it should offer significant gains over traditional methods in multi-parameter inference when the feature space dimensionality is $\mathcal{O}(10)$. 
This capability could be crucial for LHC analyses seeking to constrain multiple effective field theory operators, as was proposed for classifier-based SBI methods \cite{Brehmer:2018eca,Brehmer:2018kdj}. 
Many of the flagship LHC measurements in coming years will feature the narrow resonances, such at the Higgs, and require data-driven backgrounds, and are thus amenable to the \HISIGMA. 

A key question for future work would be to assess the impact of the dimensionality of the feature space on the accuracy of the \HISIGMA approach. 
Further validations of the accuracy of the density estimation, more rigorous uncertainty quantification methods on the density estimates, and full coverage studies would be needed prior to deployment.
For scaling to very large dimensionalities it may be interesting to explore diffusion models, which have been shown to achieve high dimensional density estimates on $O(100)$ dimensional particle physics data \cite{Mikuni:2023tok}. 
If the \HISIGMA approach is seen to not be able to accurately model very large dimensionalities, it could be used in conjunction with a classifier in a hybrid approach. 
A selection using a classifier trained on the whole feature space could be applied, and then \HISIGMA used to fit the most important features on the remaining events. 
Further work will also be needed to appropriately assess the uncertainties on this method of background estimation.

\section*{Data and Code Availability}
The datasets, and the Madgraph cards used to generate them, are provided on \href{https://doi.org/10.5281/zenodo.15587841}{Zenodo}.
Code used to process the data, train the models and evaluate them is available on \href{https://github.com/OzAmram/HI-SIGMA/tree/main}{Github}. 

\section*{Acknowledgments}
We thank Aishik Ghosh for useful discussions, particularly on parameterizing uncertainties arising from finite training statistics. We thank  Nicholas Smith  and Prasanth Shyamsundar for useful discussions and probing questions. We also thank Adrien Bayer and Caroline Cuesta for useful discussions in the early development of this project, as well as Prasanth Shyamsundar and Nick Smith. 
This work was initiated at the Aspen Center for Physics, supported by National Science Foundation grant PHY-2210452.
OA is supported by FermiForward Discovery Group, LLC under Contract No. 89243024CSC000002 with the U.S. Department of Energy, Office of Science, Office of High Energy Physics.
MS acknowledges support in part by the DOE grant de-sc0011784 and NSF  OAC-2103889, OAC-2411215, and OAC-2417682.


\bibliography{main}

\end{document}